\documentclass[preprint,superscriptaddress,longbibliography]{revtex4-1}
\usepackage{graphicx}
\usepackage{epstopdf, epsfig}
\usepackage{hyperref}

\usepackage{amssymb}
\usepackage{subfigure}
\usepackage{amsmath}

\usepackage{graphicx}
\usepackage{epstopdf}
\usepackage{color}
\usepackage{ulem}

\begin{document}

\title{Parametric control of self-sustained and self-modulated optomechanical oscillations}

\author{Stamatis Christou}
\affiliation{School of Applied Mathematical and Physical Sciences, National Technical University of Athens, Athens 15780, Greece}

\author{Vassilios Kovanis}
\affiliation{Bradley Department of Electrical and Computer Engineering, Virginia Tech Research Center in Arlington Virginia, Arlington Virginia 22203, USA}

\author{Antonios E. Giannakopoulos}
\affiliation{School of Applied Mathematical and Physical Sciences, National Technical University of Athens, Athens 15780, Greece}

\author{Yannis Kominis}
\email[corresponding author (email): ]{gkomin@central.ntua.gr}
\affiliation{School of Applied Mathematical and Physical Sciences, National Technical University of Athens, Athens 15780, Greece}

\begin{abstract}
Optomechanical systems are known to exhibit a rich set of complex dynamical features including various types of chaotic behavior and multi-stability. Although this exotic behavior has attracted an intense research interest, the utilization of optomechanical systems in technological applications, in most cases necessitates a complex, yet predictable and controllable, oscillatory response. In fact, the various types of robust oscillations supported by optomechanical systems are nested in either the same or neighboring regions of the parameter space, where chaos exists. In this work we systematically dissect the parameter space of the fundamental optomechanical oscillator in order to identify regions where stable self-sustained and self-modulated oscillations exist, by utilizing bifurcation analysis and advanced numerical continuation techniques. Moreover,in cases where bistability occurs, we study the accessibility of these oscillatory states in terms of initial conditions and their location with respect to well-defined basins of attraction. The results provide specific knowledge for the parameter sets enabling the appropriate oscillatory response for different types of applications. 
\end{abstract}

\maketitle

\section{Introduction}
Cavity optomechanics is a research field of intense interest, where the interaction between electromagnetic radiation and small-scale mechanical motion is studied as a fundamental physical process as well as a mechanism which is useful for various applications \cite{RMP_14, Metcalfe_14}. These applications include, among others, optomechanical cooling \cite{Elste_09}, highly sensitive optical detection and manipulation of small forces, displacements, masses and accelerations \cite{Painter_12, Yang_16a}, topological energy transfer \cite{Harris_16}, creation of nonclassical light-matter states for quantum information processing \cite{Reed_17}, and phonon lasing \cite{Vahala_09, Yang_18}. Such applications have been implemented and experimentally studied in a large variety of optomechanical platforms with characteristic masses, frequencies and other parameters spanning several orders of magnitude \cite{RMP_14}. It is also worth mentioning relevant applications in cavity-free optomechanical systems implemented with the utilization of dual-nanoweb optical fibers and nanostructured waveguides \cite{Russell_14, Russell_16, Russell_17, Sukhorukov_18}.

The underlying dynamics enabling the aforementioned applications can be described with coupled quantum Langevin equations for the cavity field and the mechanical amplitudes. However, for sufficiently large photon and phonon numbers, an averaged classical model for the electric field and the oscillation position can be utilized. The latter consists of a fully nonlinear system of coupled equations, supporting a rich set of complex nonlinear dynamical features, which are of crucial importance to the various applications. This set includes the bistability of steady states \cite{Dorsel_83, Mancini_94}, the existence of self-induced oscillations corresponding to stable limit cycles \cite{Vahala_05a, Vahala_05b,  Marquardt_08a, Marquardt_08b, Marquardt_14, Painter_15, Buters_15} and related multistability effects \cite{Marquardt_06, Fehske_16a} as well as the existence of self-modulated oscillations corresponding either to stable limit tori or chaotic attractors \cite{Vahala_07, Marino_11, Grebogi_14, Yang_16b, Liu_17, Marquardt_20}. Characteristic features of classical chaos such as period-doubling \cite{Feshke_15, Wang_16, Urrios_17} and quasi-periodic \cite{Marquardt_20} route to chaos, transient and intermittent chaos \cite{Lu_20}, and antimonotonicity \cite{Kingni_20}, have also been reported. Moreover, complex collective dynamics and synchronization effects have been studied in large optomechanical arrays \cite{Marquardt_11, Zhang_12, Milburn_12, Marquardt_13,  Marquardt_16} and dimer systems \cite{ElGanainy_16, Fehske_16b, Alu_17, Nori_19, Vitali_20}. It is worth mentioning that 
such dynamical features are also common to configurations where a cavity field is coupled to another optical field either in a master-slave \cite{Kovanis_95a, Kovanis_95b, Kovanis_97} or in a mutual \cite{Kominis_17, Kominis_20} fashion.

Although many theoretical works have been mainly focused on the exotic features of chaotic dynamics, the existence of various types of complex, yet predictable and controllable, oscillations is of crucial importance for applications. Steady states, self-sustained and self-modulated oscillations are dynamical objects of increasing subtlety, that actually coexist with the chaotic response, either within the same or neighboring parameter regions of the system. Moreover, these stable dynamical objects may coexist even for the same parameter values, resulting in multi-stability that is manifested as a sensitivity of the response with respect to initial conditions. In that sense, a mild complexity, capable of supporting the existence of complex oscillations without losing the predictability of the system's response, is strongly desired for applications. As in all nonlinear systems, the dynamical complexity increases with the injected power, resulting in multi-stability and chaoticity that render the system essentially unpredictable. However, a strong intrinsic optical and/or mechanical dissipation results in increasing levels of injected power for the existence of self-sustained oscillatory behavior. 

In this work, we systematically dissect the parameter space of the fundamental optomechanical oscillator in order to identify regions where the system supports self-sustained oscillations corresponding to stable limit cycles as well as self-modulated oscillations corresponding to quasi-period limit cycles. A detailed analysis identifies parameter regions of intermediate values for the dissipation constants and relatively low injected power that facilitate the utilization of the oscillatory features for practical applications, by raising restrictions related to requirements of extremely high quality factors and reflectivities of the mechanical devices, that depend crucially on the material properties and are important manufacturing issues.    
 
More specifically, we start from a simple investigation of the complexity of the phase space in terms of the existence and the stability of fixed points and their Hopf bifurcations giving rise to limit cycles, in order to identify value ranges for the optical and the mechanical dissipation constants where stable oscillations are supported, for relatively low values of injection power. By utilizing numerical continuation techniques \cite{Kuznetsov, MATCONT}, we obtain the dependence of the amplitude and the period of the limit cycles on the power and the detuning of the injected field, as well as their stability and bifurcations, and identify bifurcation points where stable limit tori, corresponding to quasiperiodic self-modulated oscillations, emerge through Neimark-Sacker bifurcations \cite{Kuznetsov}. Therefore, we obtain the complete picture of the parameter space regarding the existence and the stability of the dynamical attractors of interest and we can identify regions where a specific attractor, either a fixed point, a limit cycle, or a limit tori solely exist, facilitating its accessibility for large regions of initial conditions. Moreover, we show that for appropriate parameter values, multi-stability can be harnessed and actually utilized in applications, by demonstrating cases of bistability between either a steady state or a self-sustained oscillation, where the response of the system can be controllably switched depending on the relative position of its initial conditions with respect to the well defined basins of attraction of each state.

The paper is organized as follows: In Section II, the fundamental model is described along with the existence, stability and bifurcations of its steady states, and the dependence of the complexity of the phase space on the optical and mechanical dissipation constants is investigated. In Section III, the parameter space is systematically dissected with respect to the injected field power and frequency detuning, in order study the existence and stability of limit cycles corresponding to self-sustained oscillations as well as their bifurcations. The phase space is particularly studied for cases where bistability occurs and basins of attraction are identified. Moreover, limit tori bifurcations giving rise to self-modulated oscillations are pinpointed in the parameter space and cases of quasi-periodically and chaotically modulated oscillations are shown. The conclusions are summarized in Section IV.

\section{Fundamental Model and Steady States}
The dynamics of the fundamental optomechanical system, consisting of a single optical and a single mechanical mode, is governed by the following equations \cite{RMP_14}
\begin{eqnarray}
 \frac{d\tilde{a}}{dt}&=&\left[i\left(\Delta+x\right)-\frac{\kappa}{2}\right]\tilde{a}+\frac{1}{2} \nonumber \\
 \frac{d^2 x}{dt^2}&=&-x+P|\tilde{a}|^2-\gamma\frac{dx}{dt} \label{system}
\end{eqnarray}
where $\tilde{a}$ is the normalized complex amplitude of the electric field of the cavity mode and $x$ is the dimensionless position of the mechanical oscillator. The optical and the mechanical modes have frequencies $\omega_C$ and $\omega_M$, respectively. Time $t$ is multiplied by $\omega_M$ and the dissipation constants of the optical and the mechanical mode, $\kappa$ and $\gamma$, are also normalized to $\omega_M$. The optical mode is pumped by an external laser field with complex amplitude $\tilde{a}_L$ and frequency $\omega_L$, and $\Delta=(\omega_L-\omega_C)/\omega_M$ measures the normalized relative detuning between the pump and the cavity mode. The parameter $P=8|\tilde{a}_L|^2 g^2/\omega_M^4$ corresponds to the dimensionless pump laser power including also the strength of the optomechanical interaction through the optomechanical coupling constant $g$.
The parameter space of the system is four-dimensional $(P,\Delta,\kappa,\gamma)$ and the essential dynamical features depend strongly on the parameter range of operation. In the following, we systematically investigate the existence and the stability of dynamical attractors such as fixed points, limit cycles and limit tori, corresponding respectively to steady states, self-sustained and self-modulated oscillations. 

\subsection*{Bifurcations of Fixed Points}
The fixed points of the system are given by taking the time derivatives equal to zero, in Eq. (\ref{system}). The resulting solutions of the system of algebraic equations are given in terms of the roots of the third-order equation for the position $x$
\begin{equation}
 x^3+2\Delta x^2+\left(\Delta^2+\frac{\kappa^2}{4}\right)x-\frac{P}{4}=0
\end{equation}
with the amplitude $\rho$ and the phase $\phi$ of the electric field $\tilde{a}=\rho e^{i\phi}$ given by $\rho^2=x/P$ and $\cos\phi=\kappa \rho$. It is worth noting that the dissipation constant of the mechanical mode $\gamma$ does not appear in the above equation suggesting that the number and the values of the fixed points do not depend on it. However, $\gamma$ appears in the Jacobian of the system (\ref{system}), the eigenvalues of which determine the linear stability of the corresponding fixed points, and plays a crucial role for the existence of the various dynamical attractors of the system.

The existence and the stability of the fixed points provide a first estimation for the complexity and the structure of the phase space of the system and its dependence on the parameters. This dependence is depicted in Fig. 1, where the number and the stability of the fixed points are given as a function of the power $(P)$ and the detuning $(\Delta)$ for various values of optical $(\kappa)$ and mechanical $(\gamma)$ dissipation constants. It is clear that higher values of the optical dissipation constant $\kappa$ results in extended regions where a single stable steady state exists and moves bifurcations, occurring at the boundaries of the different regions, at higher values of $P$. The same also holds for higher values of the mechanical dissipation constant $\gamma$, except that the bistability region expands in the $P,\Delta$ subspace and its lower boundary (cusp) does not move to higher values of $P$. It is worth emphasizing that the Hopf bifurcation curves at the corresponding region boundaries provide the parameter range of values where limit cycles, corresponding to self-sustained oscillations, exist. 

As intuitively expected, in the cases of higher dissipation, a higher level of injected power is required for the support of an interesting complex dynamical behavior of the system. From the point of view of applications it is useful to have low dissipation constants so that lower levels of injected power are required. However, this is not always possible in realistic configurations, due to quality factor limitations. The parameter investigation depicted in Fig. 1 suggests that dissipation values such as $\kappa=0.1$ and $\gamma=0.01$ enable the existence of a rich set of dynamical features for relatively low values of the injected power $P$, which is highly desirable for practical applications. Such intermediate values of $\gamma$ are 2 orders of magnitude higher than previously considered for the existence of complex dynamics \cite{Marquardt_20}, thus providing a wider range of possible applications.

\section{Self-sustained and Self-modulated oscillations}
Self-sustained oscillations correspond to stable limit cycles of the optomechanical system. In the following we focus in the study of the existence, stability and bifurcations of the limit cycles for the case of dissipation constants $\kappa=0.1$ and $\gamma=0.01$, shown in Fig. 1. For such values of dissipation constants, the limit cycles bifurcate from fixed points through Hopf bifurcations for relatively small power values. Furthermore, the limit cycles undergo other bifurcations giving rise to stable limit tori corresponding to quasi-periodic self-modulated oscillations.

\subsection*{Bifurcations of Limit Cycles and Limit Tori}
 Close to the Hopf bifurcation points the limit cycles have infinitesimal oscillation amplitudes. In order to calculate their amplitudes and periods as well as their stability changes and bifurcations, advanced numerical continuation techniques are required, such as those of the continuation toolbox MatCont \cite{MATCONT}, utilized in this work. A dissection of the parameter space with respect to $P$ results, according to Fig. 1, shows that, depending on the power level $P$ and its relative position with respect to the local extrema of the Hopf bifurcation curve, for varying detuning $\Delta$ we can have either two or four Hopf points, resulting to either one or two families of limit cycles.

Figure 2, depicts the projection of the limit cycles in the position-velocity $(x,v)$ plane of the phase space as well as their periods, as a function of the detuning for two power levels below the local maximum of the Hopf curve, Fig. 1. For the case (A), there exist two Hopf points in the positive detuning region, corresponding to a single family of limit cycles. Limit cycles bifurcating from the Hopf point with larger detuning $\Delta_M$ are stable whereas those bifurcating from the Hopf point with smaller detuning $\Delta_m$ are unstable. It is clear that stable and unstable limit cycles coexist for $\Delta<\Delta_m$, as they bifurcate from a Saddle-Node bifurcation of limit cycles (Limit Point Cycle - LPC). Very close to the LPC point there also exists a Neimark-Sacker (NS) bifurcation of a limit tori, which will be further studied in the following. For the case (B), there exist four Hopf points, corresponding to two families of limit cycles. The one located in the negative detuning region consists solely of stable limit cycles. However, the family of limit cycles in the positive detuning region has two Saddle-Node bifurcations of limit cycles, resulting in the coexistence of one unstable and two stable limit cycles for some range of detuning values. In this range bistability of limit cycles occurs, which will also be further studied in the following.

Cases of higher power values are depicted in Fig. 3. For such power levels the two families of the case (B) have merged to a single family of limit cycle with varying stability type for different values of the detuning. A stable limit cycle of small amplitude bifurcates from a Hopf point for large negative detuning that loses its stability through a Period Doubling (PD) bifurcation as $\Delta$ increases, as shown in case (C). The same also holds for cases (D) and (E), but is not depicted in order to focus in the region around $\Delta=0$ (and mostly $\Delta>0$) where complex bifurcations take place. In this region the family of limit cycles undergoes an increasing number of foldings for higher values of $P$ resulting in the coexistence of multiple stable and unstable limit cycles. A sequence of various bifurcations takes place including Saddle-Node bifurcations of limit cycles (LPC), Period-Doubling (PD) and Neimark-Sacker (NS) bifurcations. 

The numerical continuation of these bifurcation points for a continuous range of power ($P$) values allows for the construction of a bifurcation diagram in the $(\Delta, P)$ parameter plane as shown in Fig. 4. The locus of Neimark-Sacker bifurcations consists of smooth curves forming tongues, which are also related to tongues of Saddle-Node curves; the latter also form various cusps. Period-Doubling curves have a more complex topology, suggesting a Period-Doubling route to chaos for increasing power $P$.

\subsection*{Bistability and Basins of Attraction}
Multistability may occur under the coexistence of more than one stable attractors of the same or different type. For high power values the coexistence of multiple stable attractors along with the complexity of their basins of attraction, sometimes having a fractal-like topology, results in an exotic dynamical behavior \cite{Marquardt_20}, characterized by such a sensitivity with respect to initial conditions that essentially makes the system's response unpredictable. However, under conditions where a few simple stable attractors coexist, the phase space of the system is partitioned into a set of well-defined basins of attraction for each attractor. Close to the sharp boundaries separating different basins of attraction, the system's response is highly sensitive with respect to initial conditions as well as parameter values, resulting to a controlled sensitivity that can be very useful for applications related to sensing and measurement \cite{Marquardt_06}.

Following this approach, we investigate the bistability either between a fixed point and a limit cycle or between two limit cycles with different oscillation amplitudes, for the case of relatively small pump power $P$, corresponding to parameter values such as in Fig. 2(B), also projected in the $x,\Delta$ plane in Fig. 5(top). For detuning values corresponding to cases (a) and (b) a stable large limit cycle coexists with a stable fixed point, whereas for cases (c) and (d) the stable fixed point has lost its stability through a Hopf bifurcation and a stable limit cycle with small amplitude has emerged. The four panels (a)-(d) depict the corresponding basins of attraction in a the two-dimensional slice of the four-dimensional phase space $x_0=x(0), a_0=Re\{\tilde{a}(0)\}$, where the other initial conditions are $v(0)=0, Im\{\tilde{a}(0)\}=0$. The basin of attraction of the large limit cycle expands at the expense of the basin of attraction of the fixed point as the detuning is increased from (a) to (b). After the Hopf bifurcation point, where the stable fixed point gives rise to a stable small limit cycle, bistability between the two limit cycles takes place. The small limit cycle inherits the basin of attraction of the fixed point as shown in (b) and (c). As the detuning is further increased, the basin of attraction of the large limit cycle continuously expands, as shown in (d), until covering the whole plane of initial conditions, after the Saddle-Node (LPC) point, where the large limit cycle is the only attractor. It is worth mentioning that these diagrams not only express the relative strength of each attractor, but also provide information on its accessibility in terms of specific initial conditions. Therefore, they show to what extent a steady state or a self-sustained optomechanical oscillation can be reached from purely mechanical $(a_0=0)$ or optical $(x_0=0)$ initial conditions. Cases where specific attractors can be reached for any value of $a_0$ or $x_0$ are also shown. The exisence of well-defined basins of attraction with sharp boundaries and the detailed knowledge of the dependence of the excitation of a specific dynamical behavior on the initial conditions can be very useful for practical sensing applications.

\subsection*{Self-modulated oscillations}
Self-modulated oscillations are more subtle dynamical objects in comparison to fixed points and limit cycles. Quasi-periodically self-modulated oscillations emerge from limit cycles through Neimark-Sacker bifurcations and form limit tori in the phase space of the system \cite{Kuznetsov}. For their practical use, they have to be, not only stable, but also accessible from a large set of initial conditions. Although, in a nonlinear system we cannot exclude the possibility that an additional attractor (such as a strange attractor) coexists with the attractor of interest, the detailed analysis of the stability and the bifurcations of fixed points and limit cycles provides ranges of parameter values where no stable fixed points or limit cycles coexist with the stable limit tori of interest. Therefore, by excluding such dominant attractors, it can be reasonably expected that the basin of attraction of the limit tori covers a significant part (if not the whole) phase space. 

Along this direction, combining information for the stability of fixed points (Fig. 1) as well as the stability of limit cycles and their Neimark-Sacker bifurcations (Fig. 3(D)), we obtain a stable limit tori which is accessible from a large set of random initial conditions, as shown in Fig. 6(top). In the same parameter region, where no stable fixed points and limit cycles exist, for a different value of the detuning, a chaotically self-modulated oscillation is supported by the system, as shown in Fig. 6(bottom).

\section{Summary and Conclusions}
The rich set of complex dynamics supported by the fundamental optomechanical oscillator has been investigated in terms of identifying robust dynamical states, such as self-sustained and self-modulated oscillations, being of particular importance for applications. The parameter space of the system has been systematically dissected with the utilization of analytical and numerical continuation methods and all the important bifurcations have been pinpointed in the parameter space of the system.

The existence and the stability of families of limit cycles as well as the dependence of their amplitude and period on the parameters of the system has been studied in detail. Regions of the parameter space where bistability between a fixed point and a limit cycle or between two different limit cycles have been identified and the phase space of the system has been investigated. The formation of well-defined basins of attraction, having sharp boundaries, has been shown and implications with respect to sensing applications have been discussed. Moreover, the more subtle dynamical states of self-modulated oscillations have been located in the parameter space of the system, through bifurcations of limit cycles. The detailed knoweledge of the stability properties of  other antagonistic dynamical states, namely fixed points and limit cycles, allowed for the selection of parameter values where no such stable states exist, therefore facilitating the accessibility of the self-modulated oscillations from a large number of phase space initial conditions, by harnessing multi-stability.

The presented systematic analysis of the parameter space of the fundamental optomechanical oscillator goes beyond the study of the exotic and chaotic dynamics of the system in the sence that it allows the identification of specific parameter values for the existence of self-sustained and self-modulated oscillatory states and the facilitation of their observation in terms of initial conditions. The latter allows the parametric control and the exploitation of these states for practical applications requiring a balance between complexity and predictability of the system's response.

\section*{Acknowledgements}
YK and AEG acknowledge stimulating discussions on nonlinear oscillators with Dr. Panos Tsopelas.

\clearpage

\begin{figure*}\centering
\begin{center}
     \subfigure
       {\includegraphics[width=0.40\columnwidth]{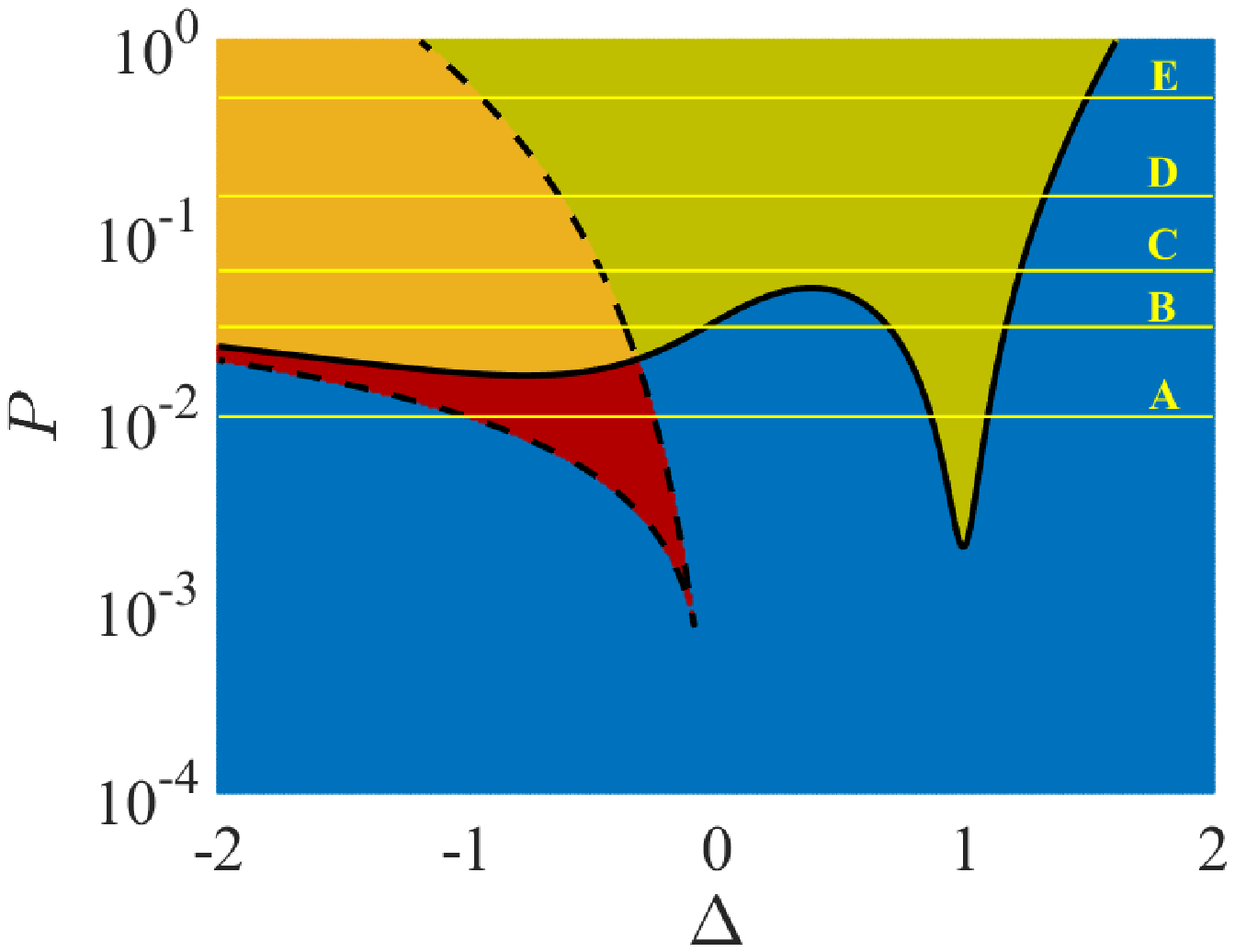}}
     \subfigure
          {\includegraphics[width=0.40\columnwidth]{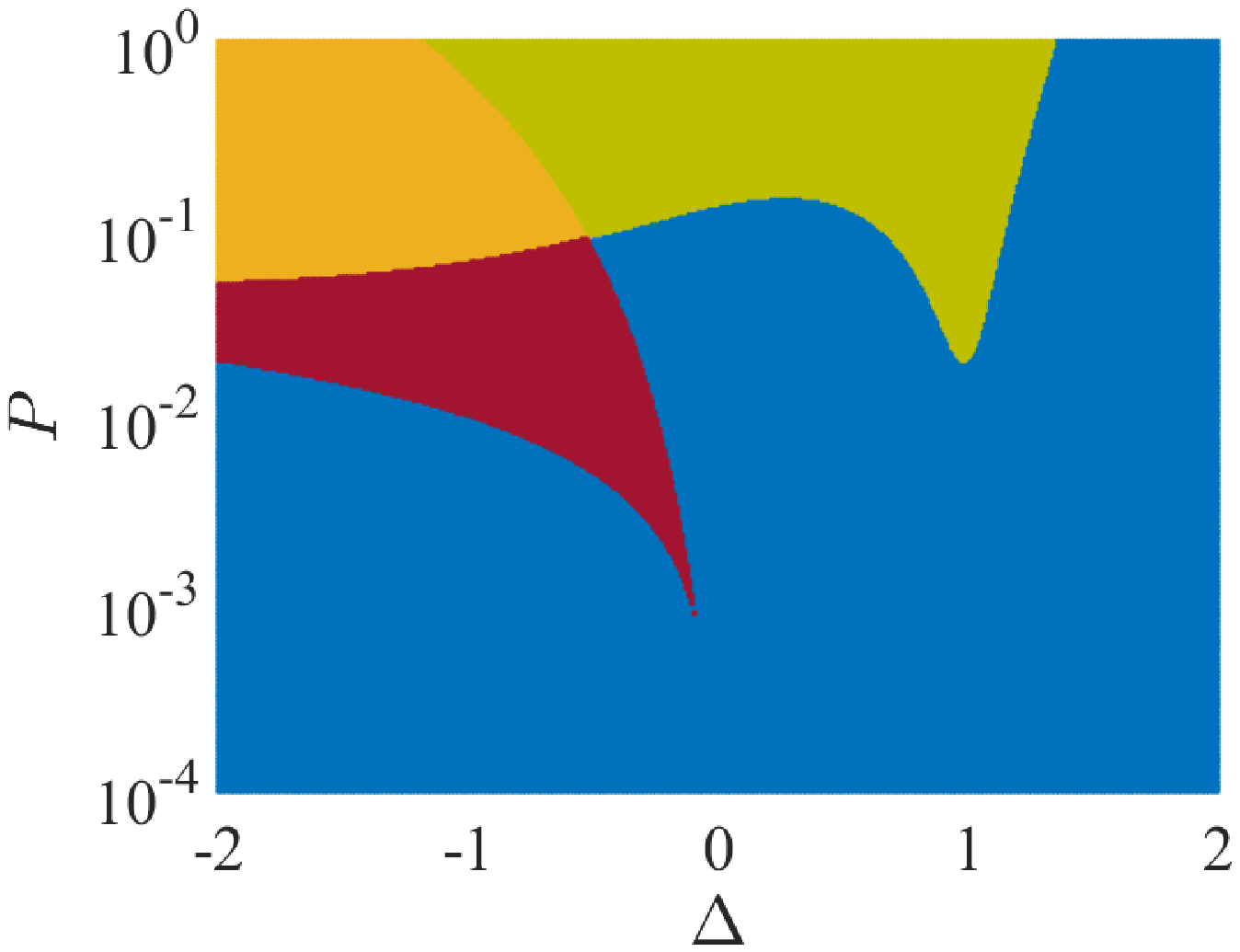}}
     \subfigure
          {\includegraphics[width=0.40\columnwidth]{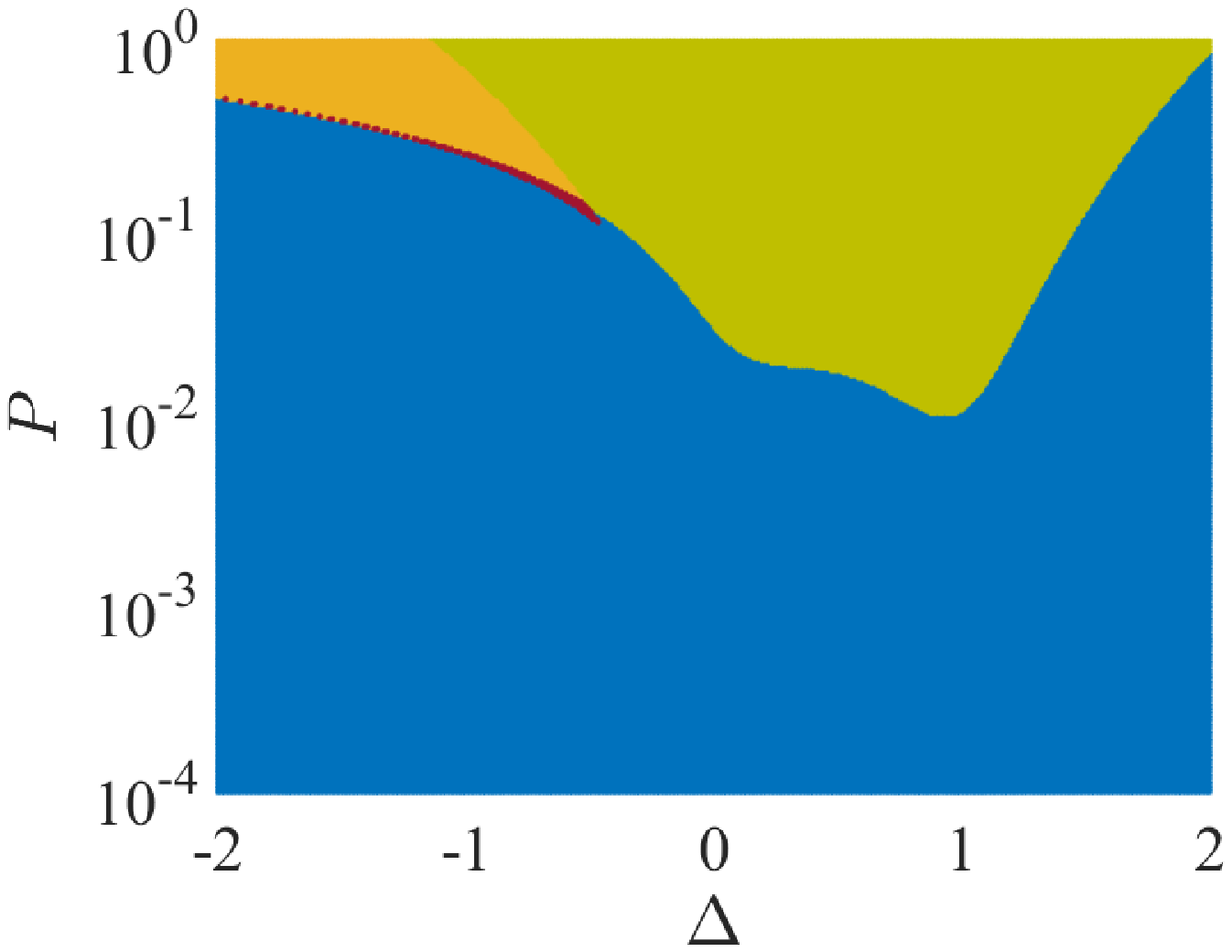}}
     \subfigure
          {\includegraphics[width=0.40\columnwidth]{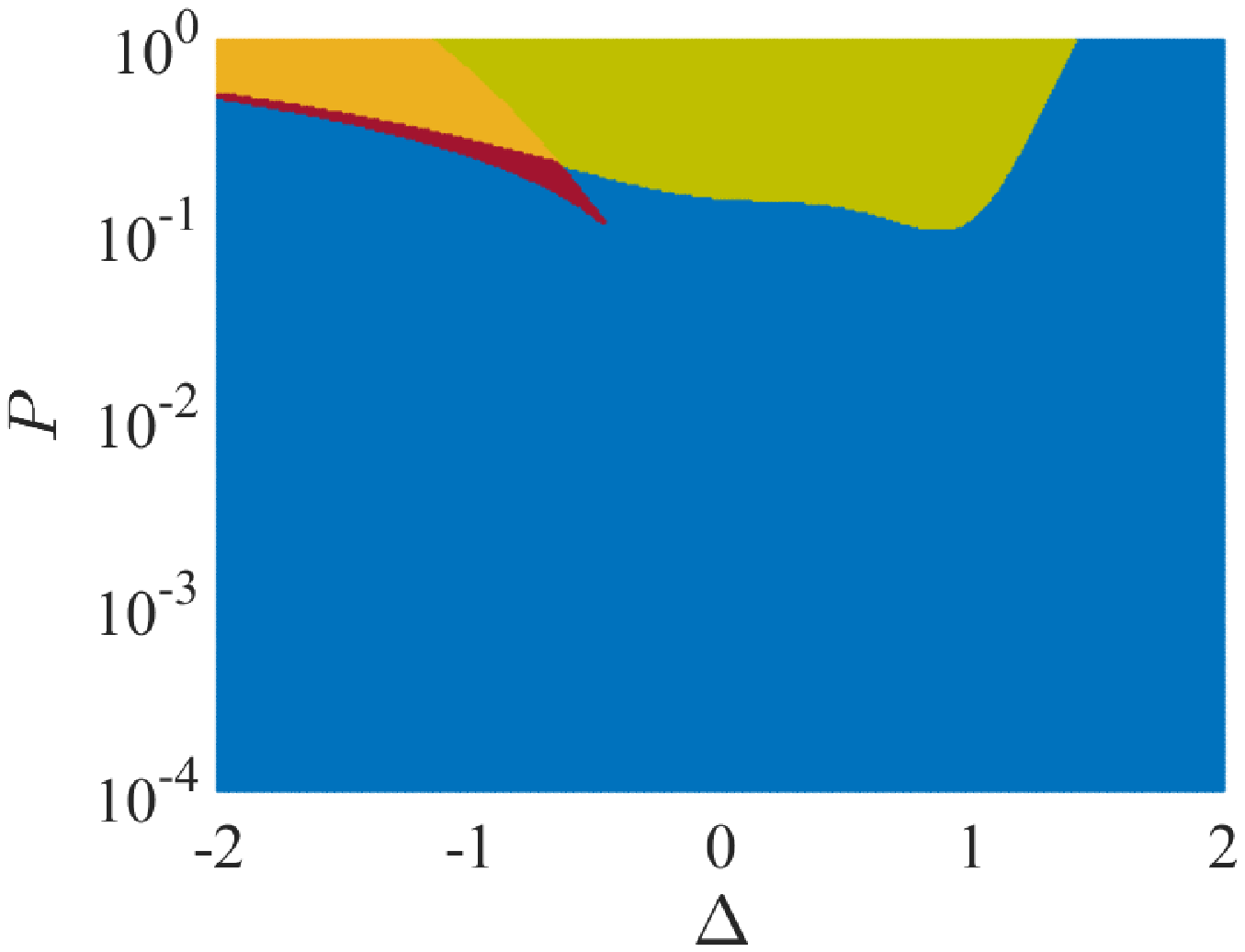}}     
     \caption{Existence and stability of fixed points in the four dimensional parameter space $(P,\Delta,\kappa,\gamma)$. The optical $(\kappa)$ and mechanical $(\gamma)$ dissipation constants are $\kappa=0.1, 0.5$ (top, bottom) and $\gamma=0.01, 0.1$ (left, right). Different colors denote different number and stability of fixed points as follows: Blue - a single stable spiral, Yellow - a stable and an unstable spiral as well as a saddle point, Red - two stable spirals (fixed point bistability) and a saddle, Green - an unstable spiral (no steady state). The dashed black line denotes a saddle-node bifurcation of fixed points and the solid black line denotes a Hopf bifurcation giving rise to limit cycles.}
\end{center}
\end{figure*}

\begin{figure*}\centering
\begin{center}
     \subfigure
       {\includegraphics[width=0.40\columnwidth]{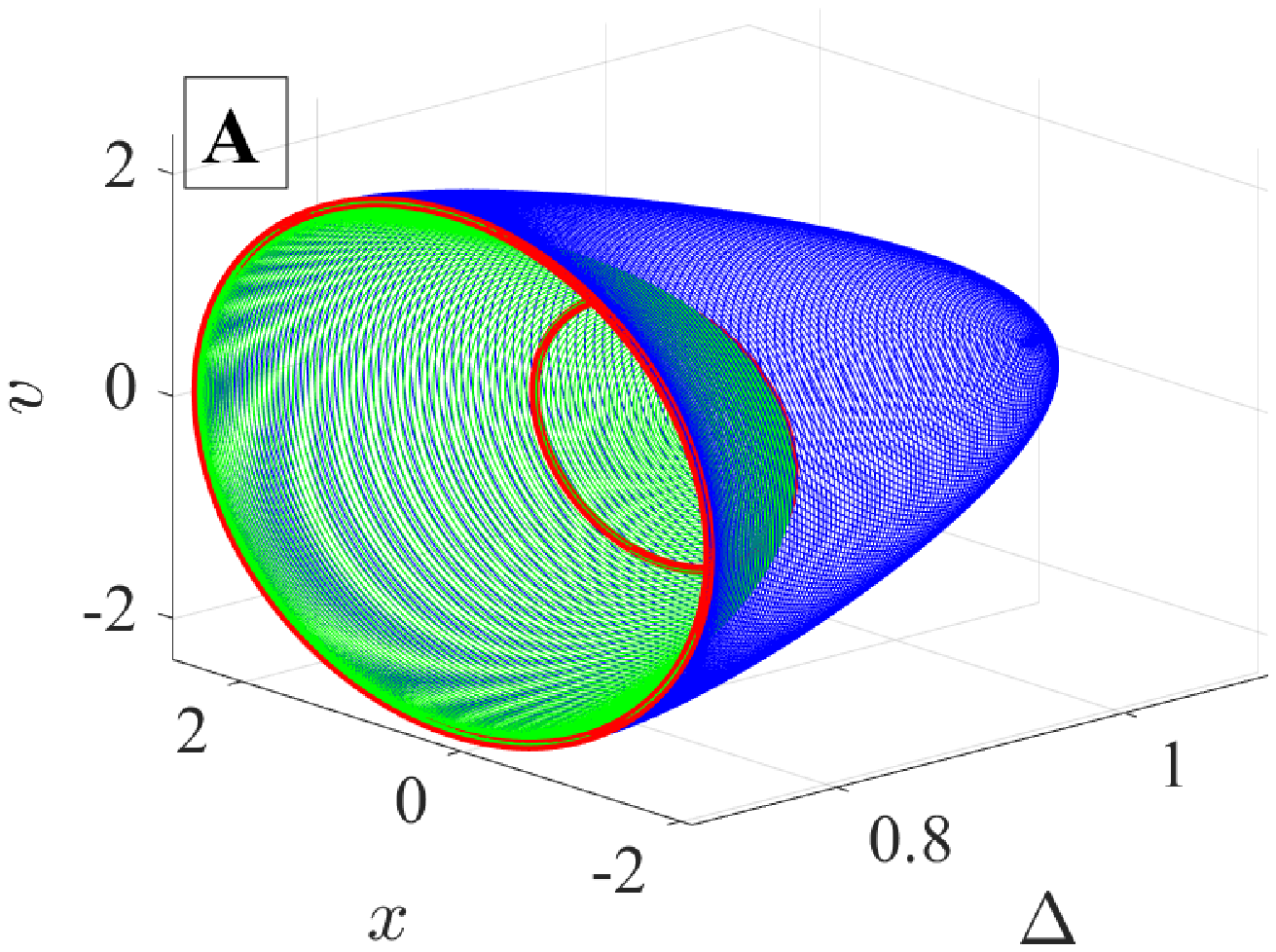}}
     \subfigure
       {\includegraphics[width=0.40\columnwidth]{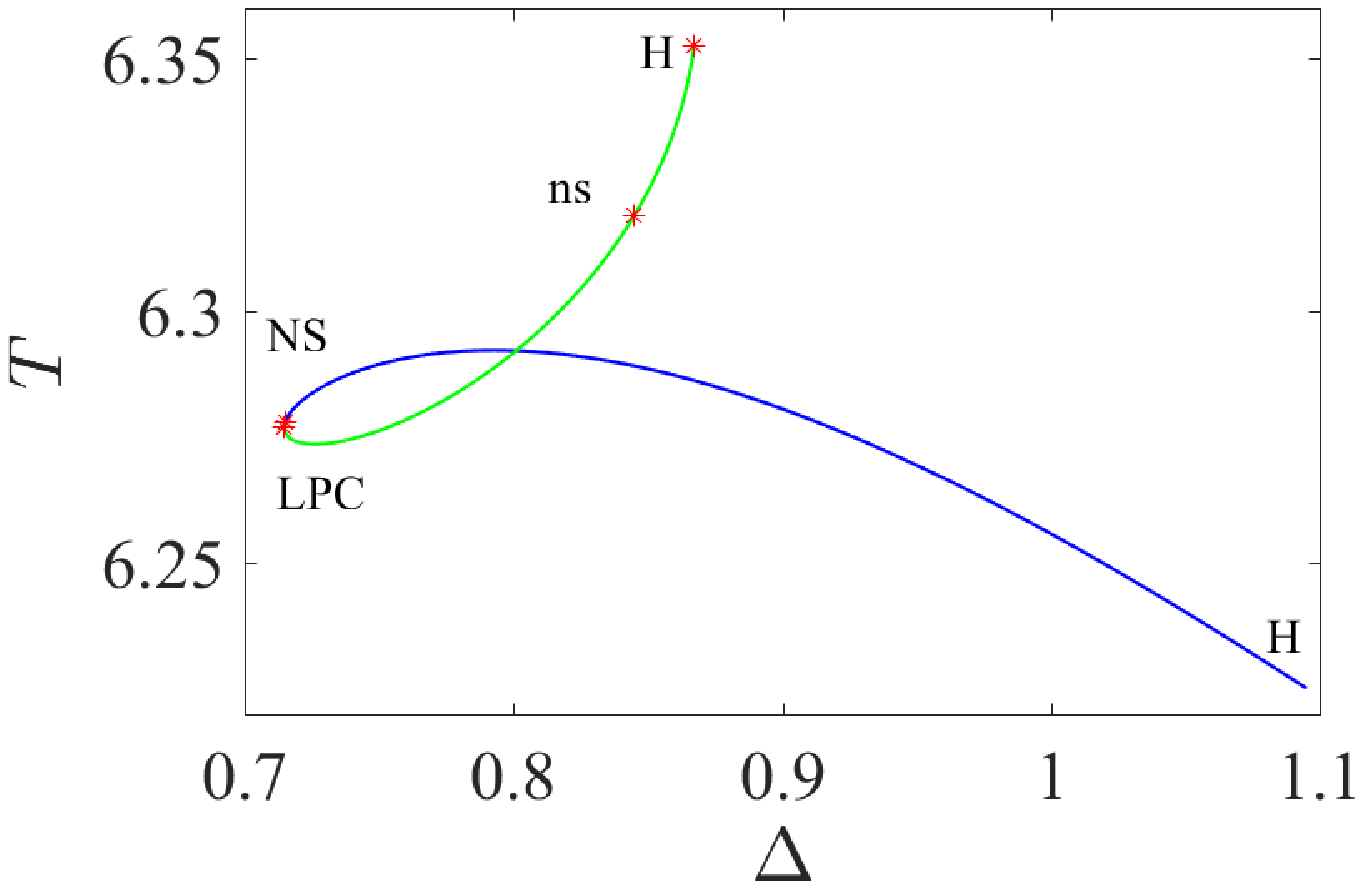}}
     \subfigure
          {\includegraphics[width=0.40\columnwidth]{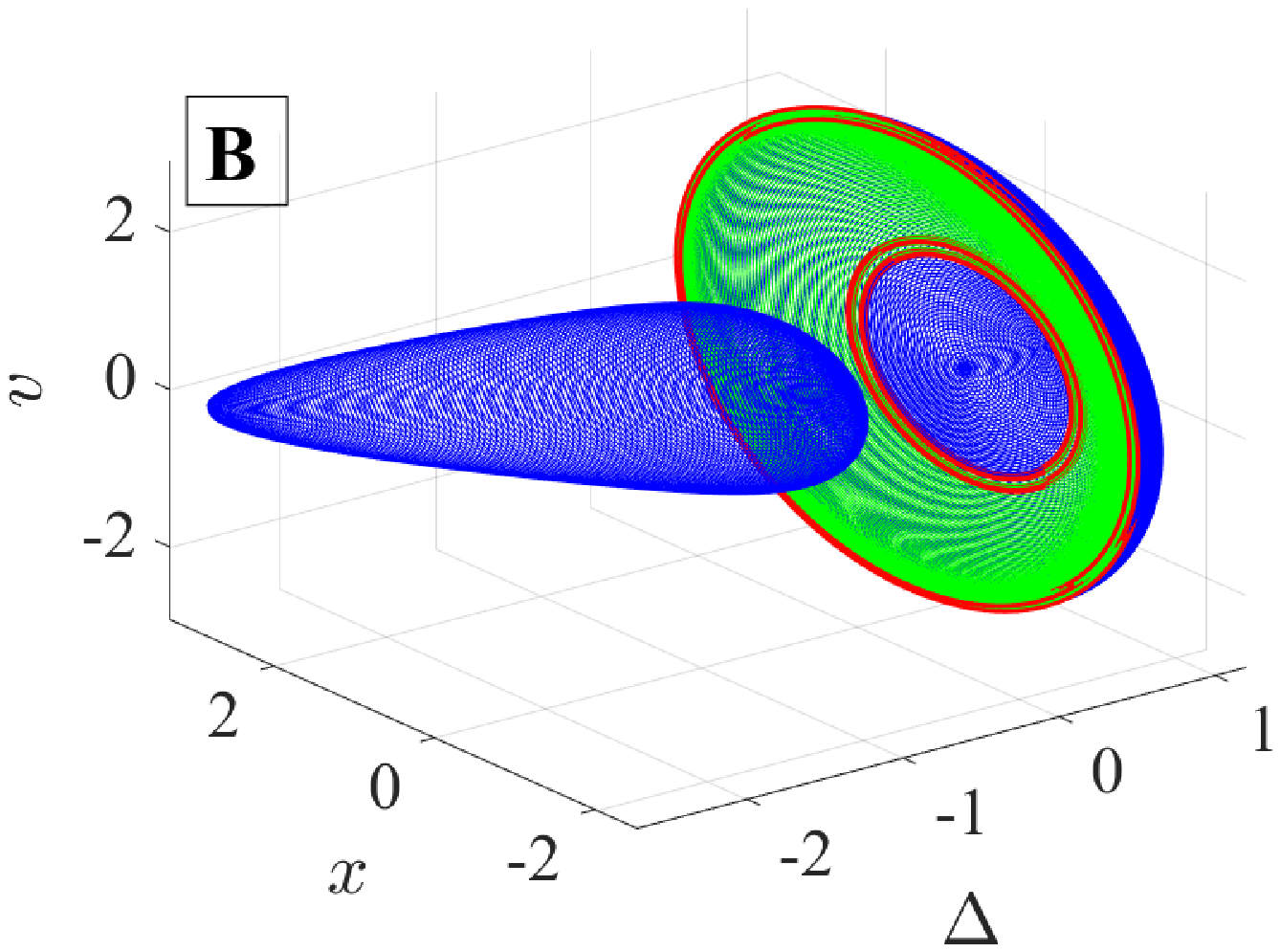}}
     \subfigure
       {\includegraphics[width=0.40\columnwidth]{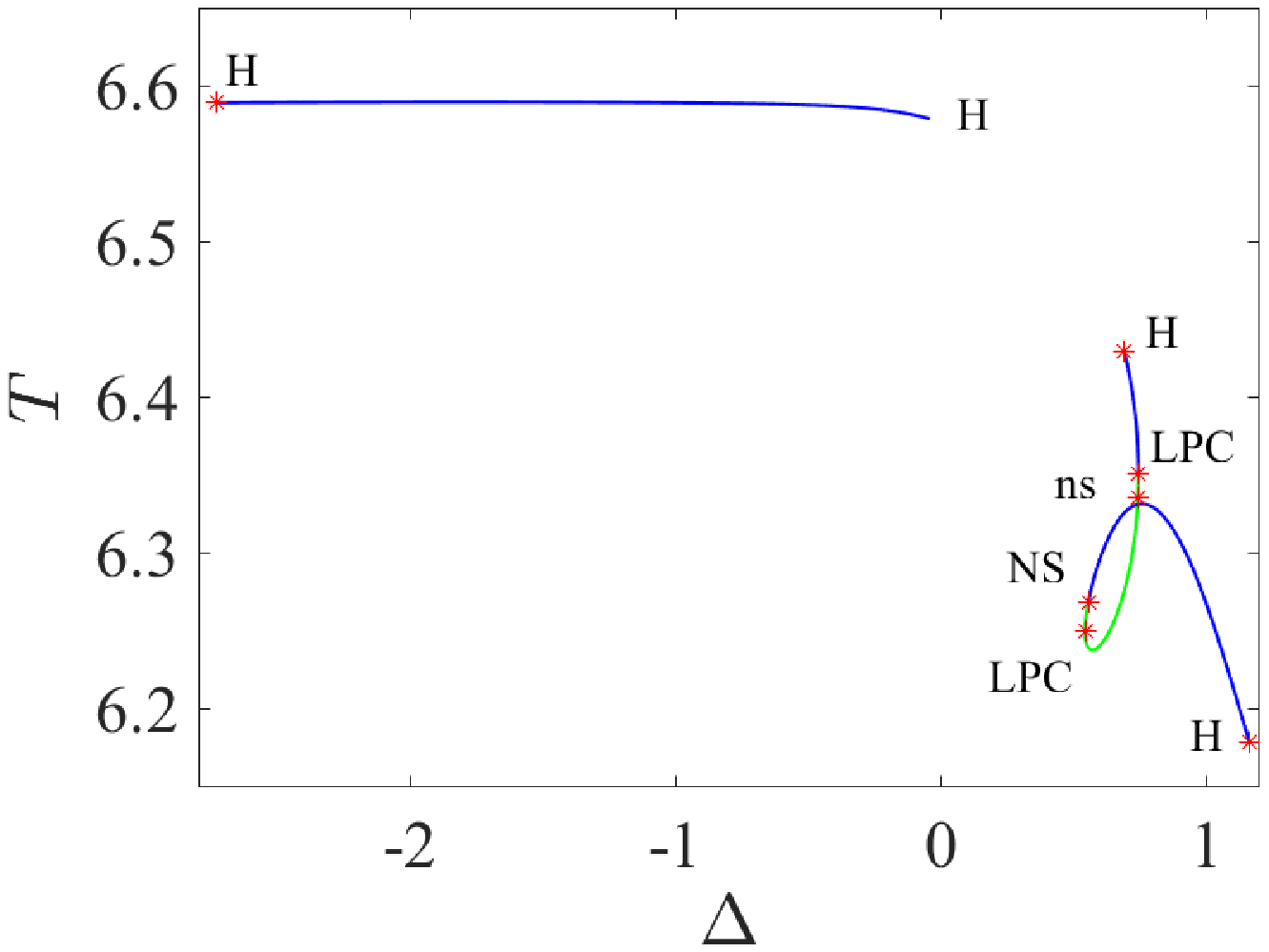}}     
     \caption{Limit cycles and their bifurcations for relatively low injection power corresponding to values shown in Fig. 1 ($\kappa=0.1$, $\gamma=0.01$). (A) $P=0.01)$, (B) $P=0.03$. Left panels: Projections of the limit cycles at the position-velocity $(x,v)$ subspace as a function of the detuning (blue - stable, green - unstable, red - bifurcation). Right panels: Period of the limit cycles as a function of the detuning. Bifurcation symbols: H - Hopf, LPC - Limit Point Cycle (Saddle-Node of Limit Cycles), NS - Neimark-Sacker (Torus), ns - neutral saddle.}
\end{center}
\end{figure*}

\begin{figure*}\centering
\begin{center}
     \subfigure
          {\includegraphics[width=0.40\columnwidth]{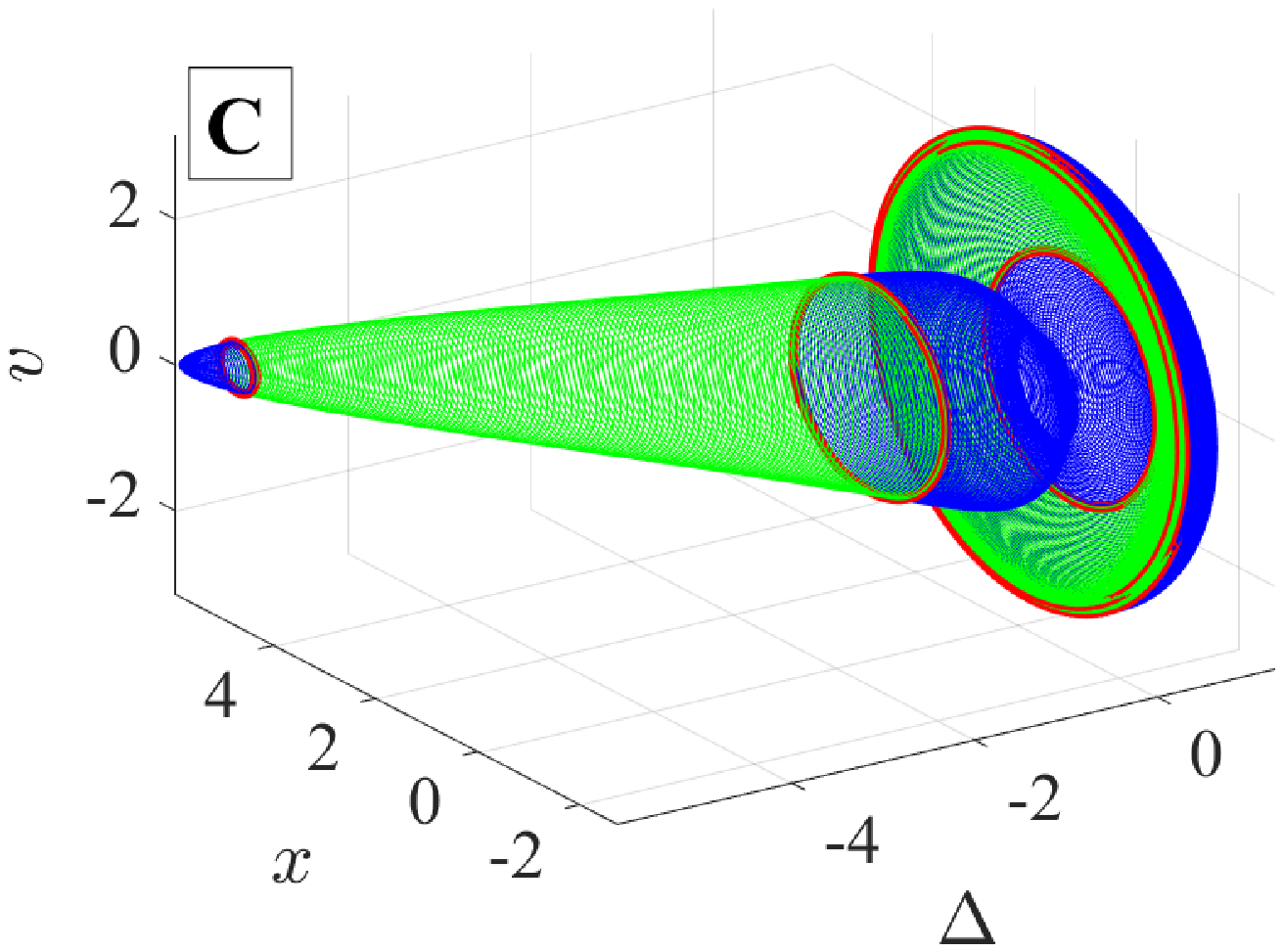}}
     \subfigure
          {\includegraphics[width=0.40\columnwidth]{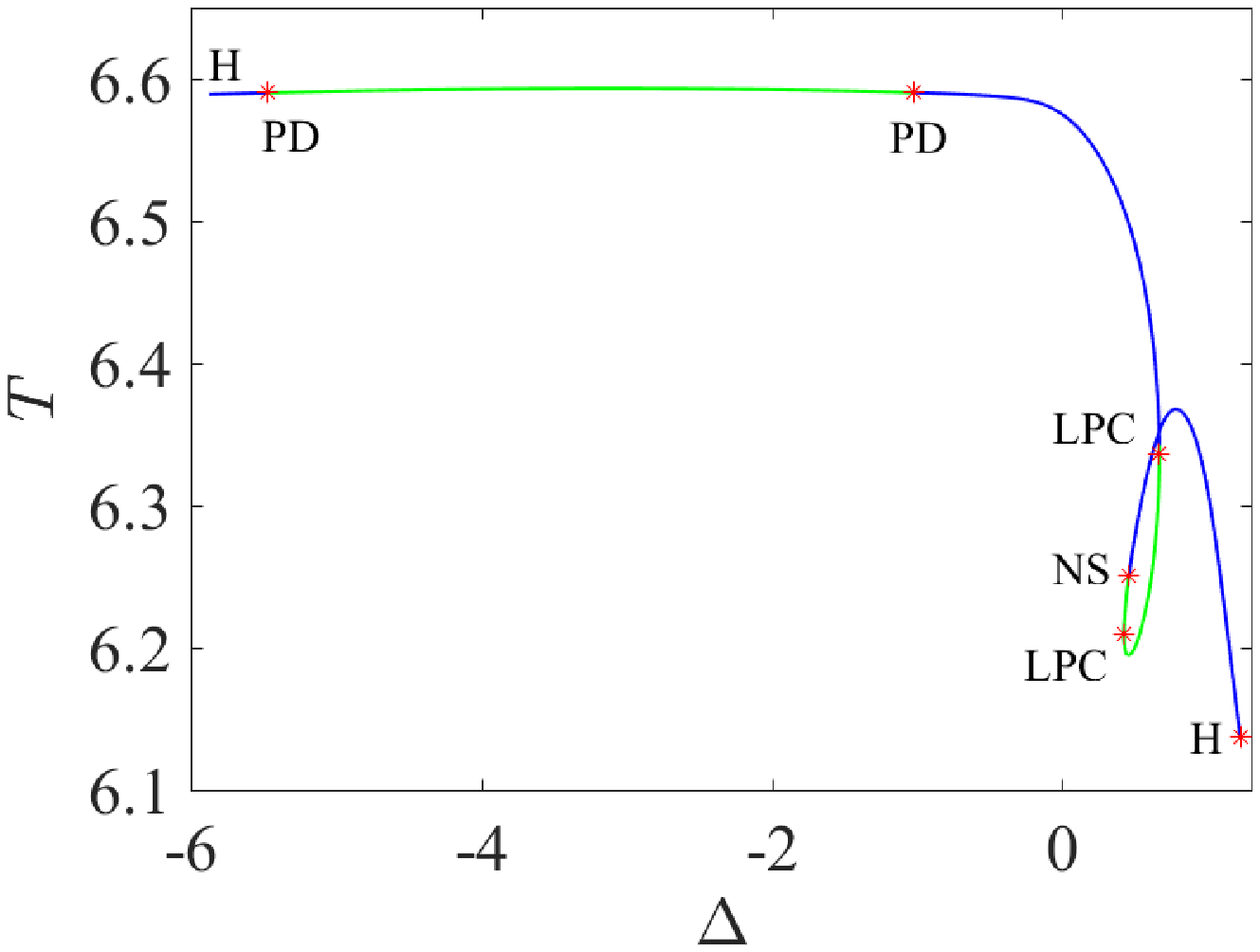}}     
     \subfigure
          {\includegraphics[width=0.40\columnwidth]{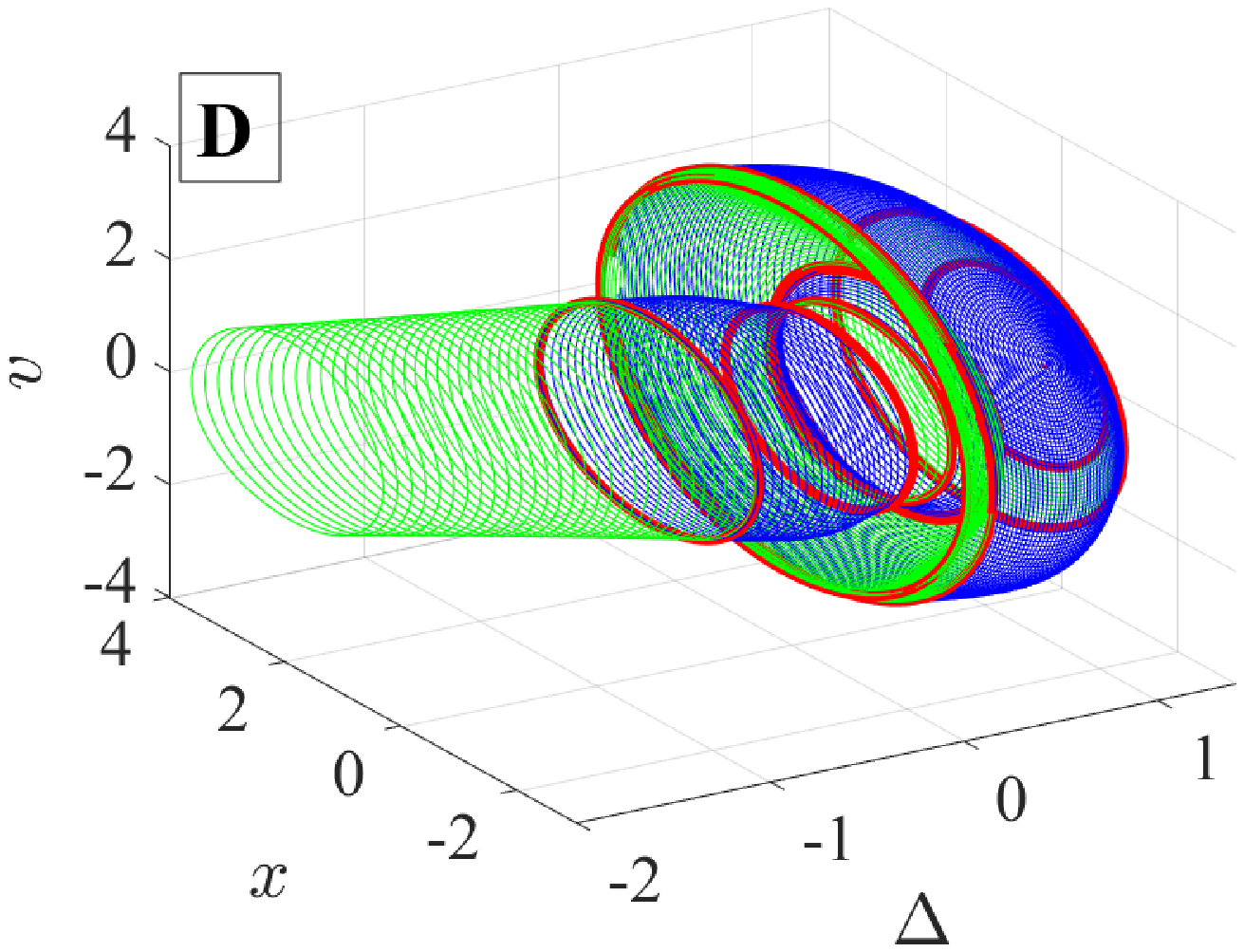}} 
     \subfigure
          {\includegraphics[width=0.40\columnwidth]{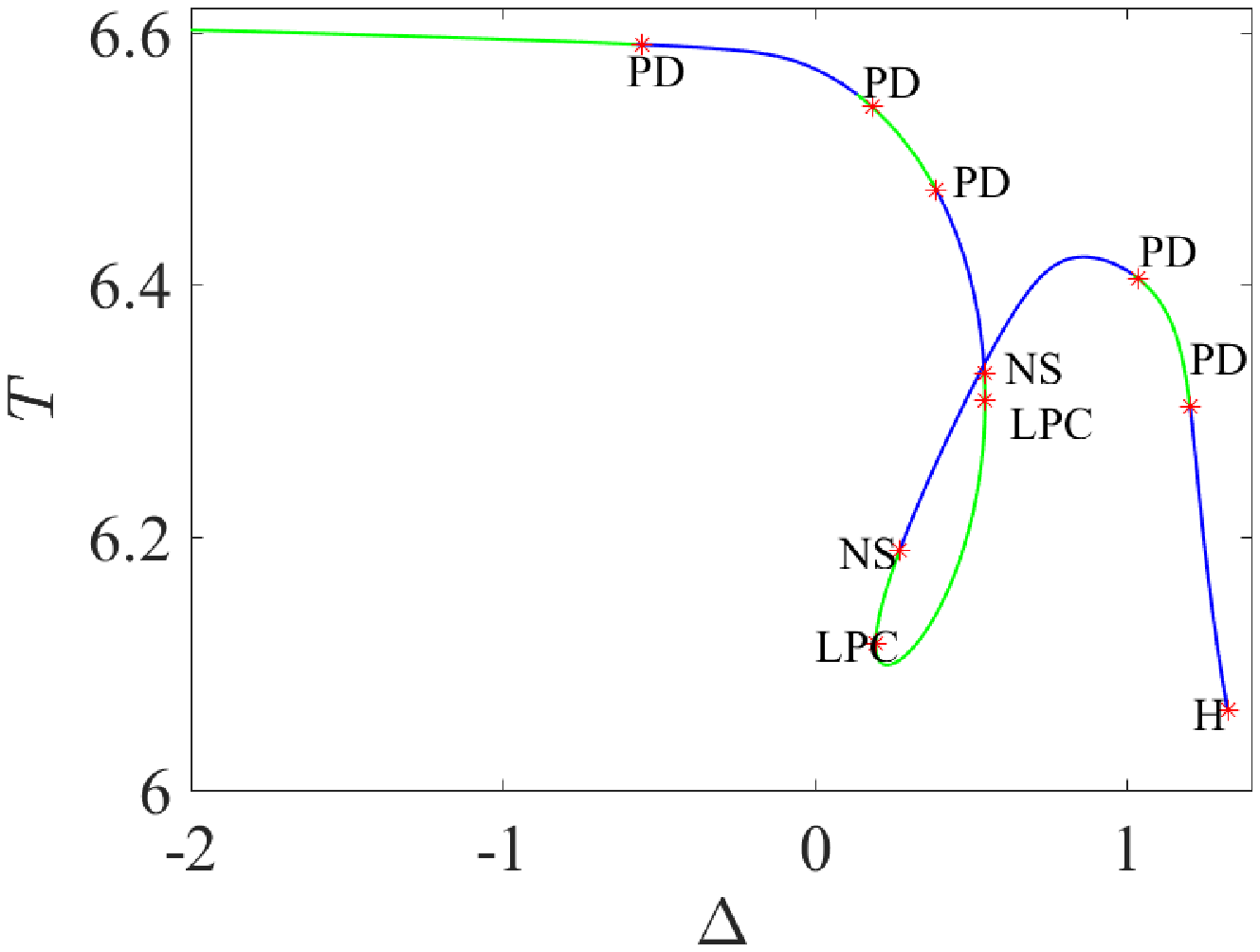}}     
     \subfigure
          {\includegraphics[width=0.40\columnwidth]{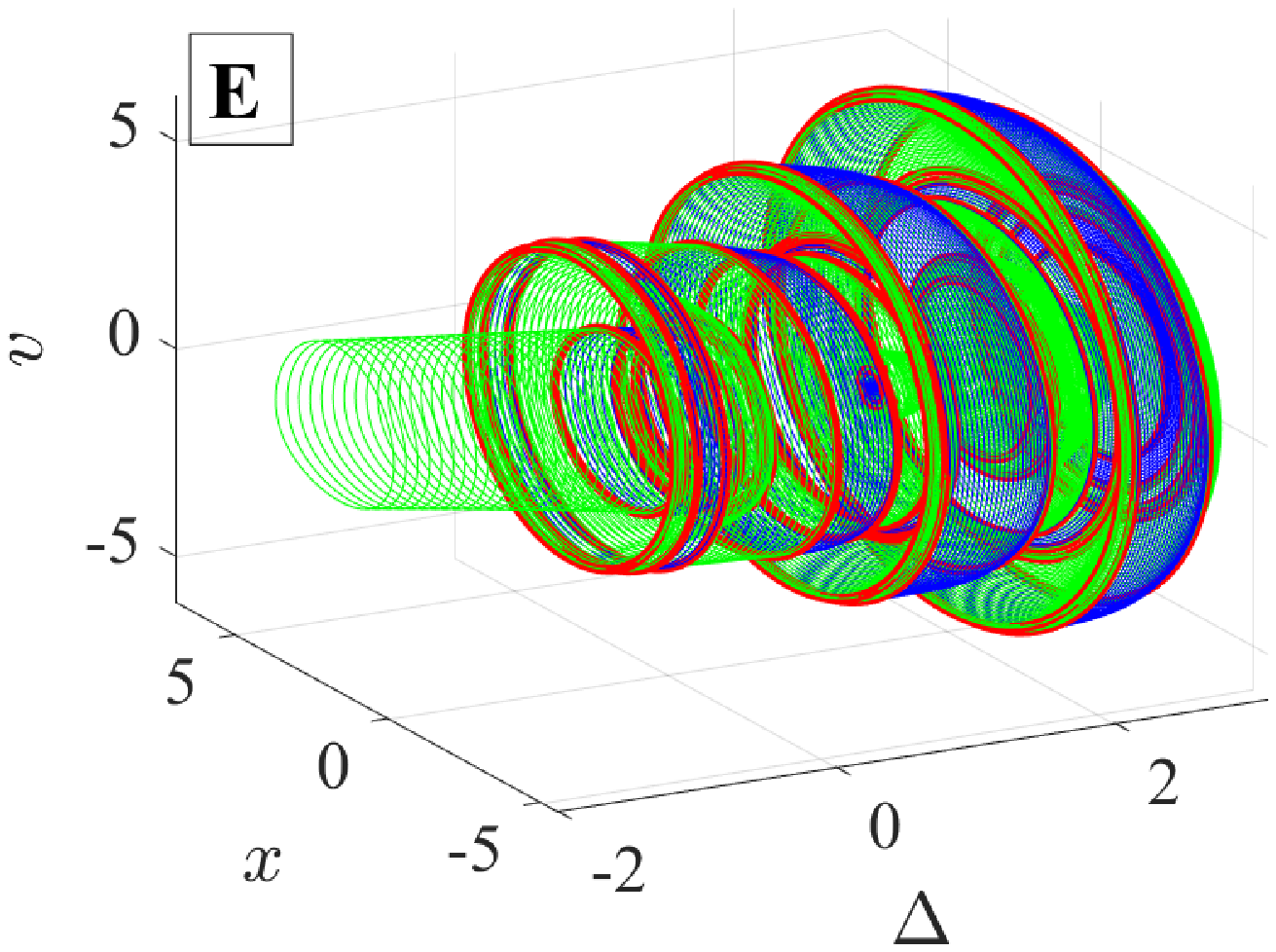}}
     \subfigure
          {\includegraphics[width=0.40\columnwidth]{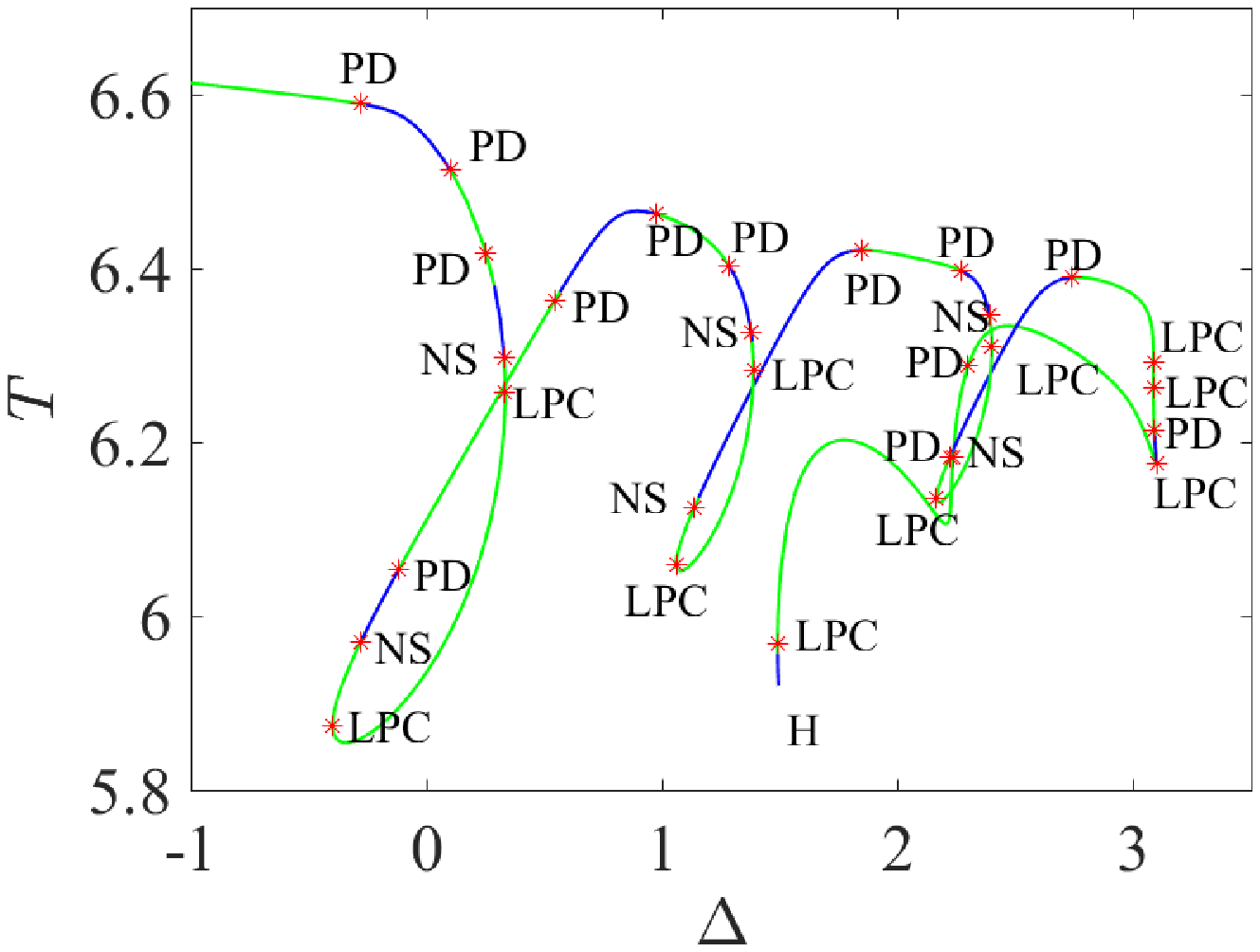}}
     \caption{Limit cycles and their bifurcations for higher injection power corresponding to values shown in Fig. 1 ($\kappa=0.1$, $\gamma=0.01$). (C) $P=0.06$, (D) $P=0.15$, (E) $P=0.50$ . Left panels: Projections of the limit cycles at the position-velocity $(x,v)$ subspace as a function of the detuning (blue - stable, green - unstable, red - bifurcation). Right panels: Period of the limit cycles as a function of the detuning. Bifurcation symbols: H - Hopf, LPC - Limit Point Cycle (Saddle-Node of Limit Cycles), NS - Neimark-Sacker (Torus), ns - neutral saddle. For the cases (D) and (E), the range of values of the detuning $\Delta$ is restricted in the neighborhood of the zero detuning, where bifurcations actually occur. For smaller (negative) values of $\Delta$ the dependence is qualitatively similar to case (C). }
\end{center}
\end{figure*}

\begin{figure*}\centering
\begin{center}
     \subfigure
          {\includegraphics[width=0.60\columnwidth]{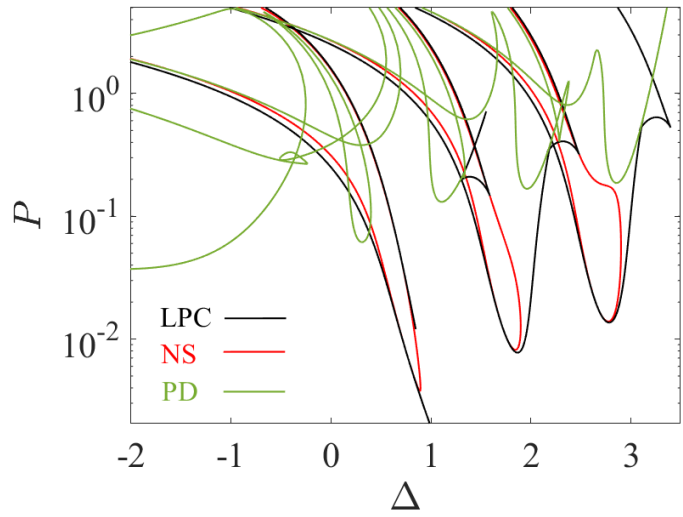}}\\
     \caption{Limit cycle bifurcations in the $P, \Delta$ parameter subspace for dissipation values $\kappa=0.1$, $\gamma=0.01$. Black lines depict the location of the Saddle-Node bifurcations of limit cycles (LPC), red lines depict the location of Neimark-Sacker (NS) or torus bifurcations, and green lines depict the location of various Period-Doubling (PD) bifurcations. }
\end{center}
\end{figure*}

\begin{figure*}\centering
\begin{center}
     \subfigure
          {\includegraphics[width=0.40\columnwidth]{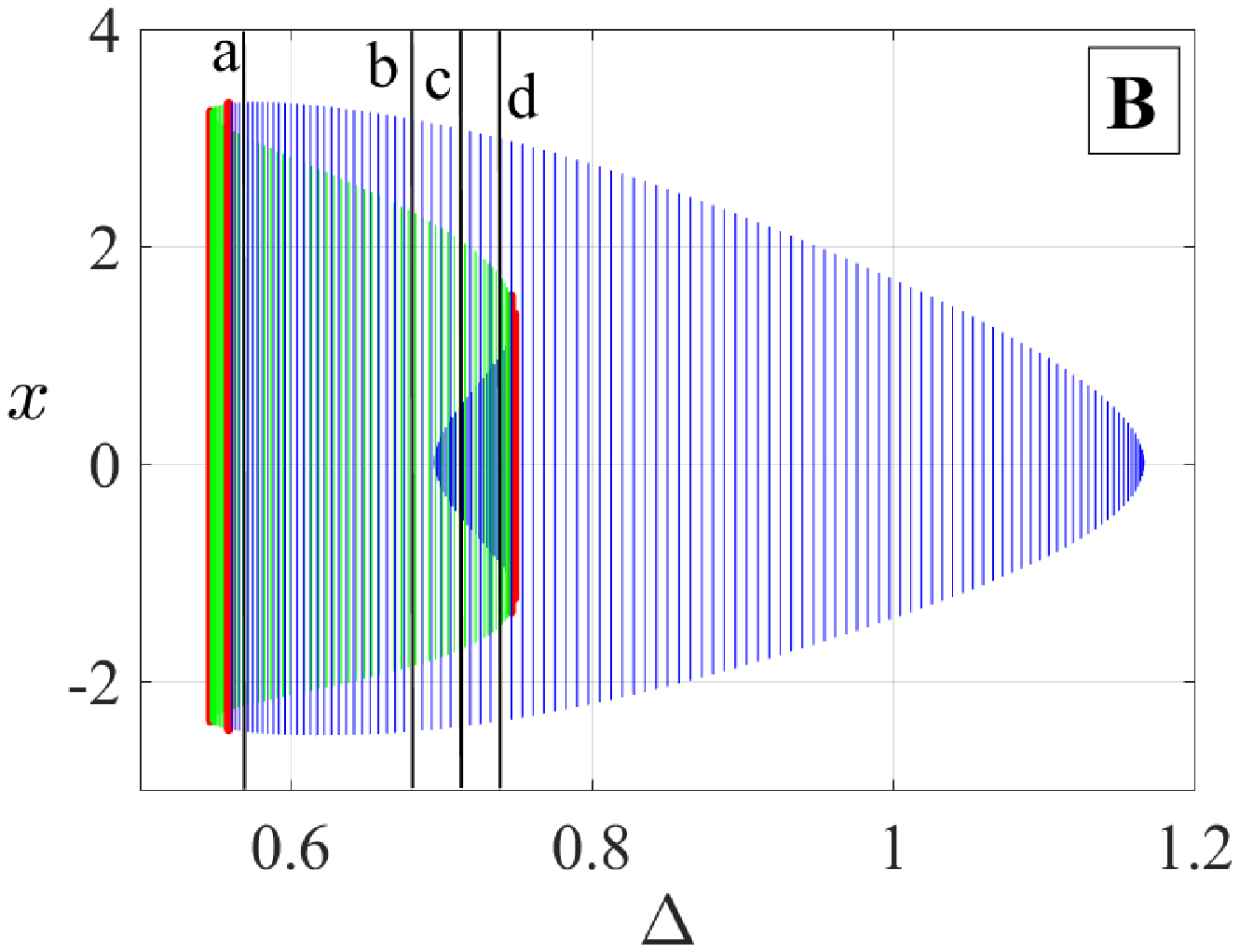}}\\
     \subfigure
          {\includegraphics[width=0.40\columnwidth]{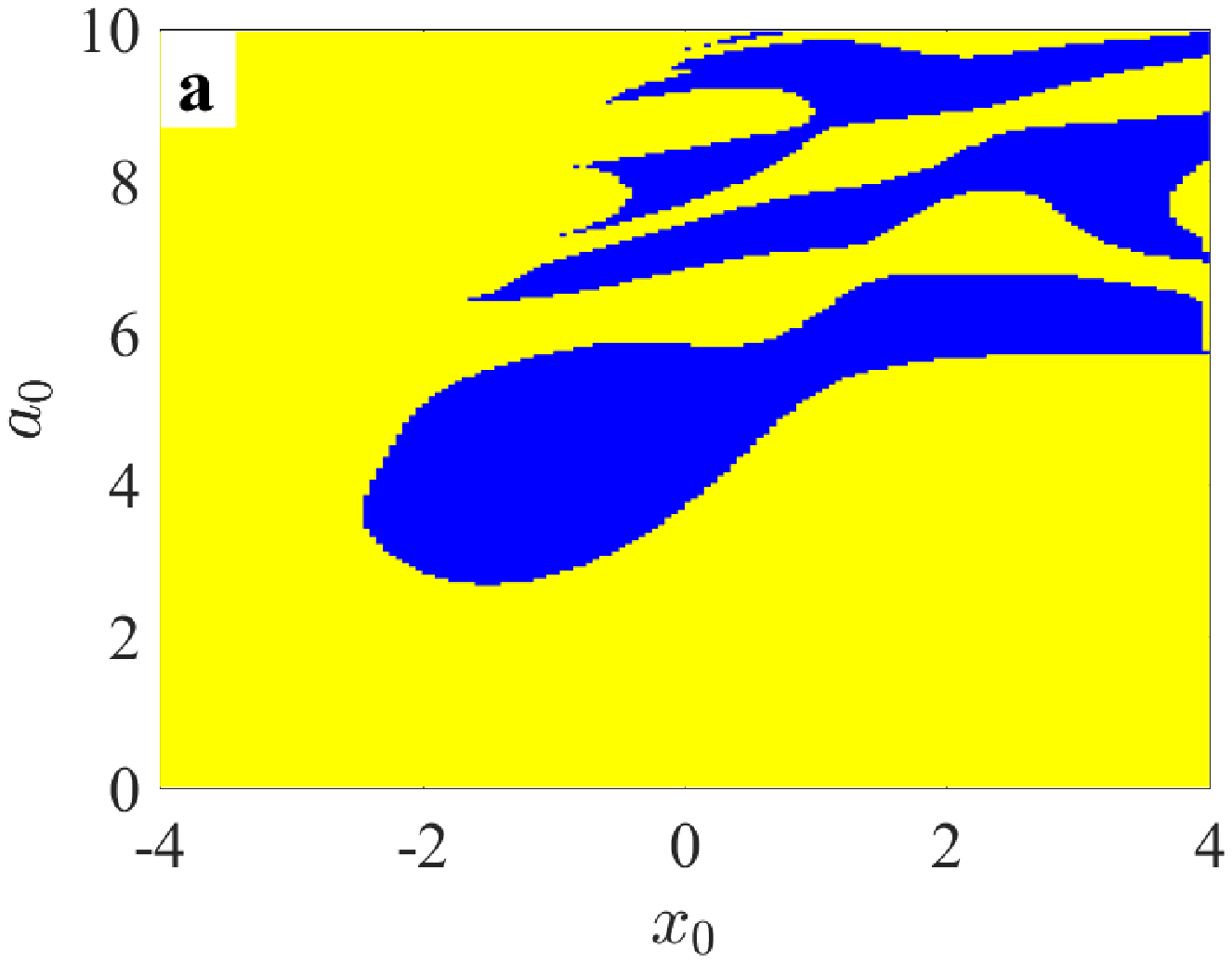}} 
     \subfigure
          {\includegraphics[width=0.40\columnwidth]{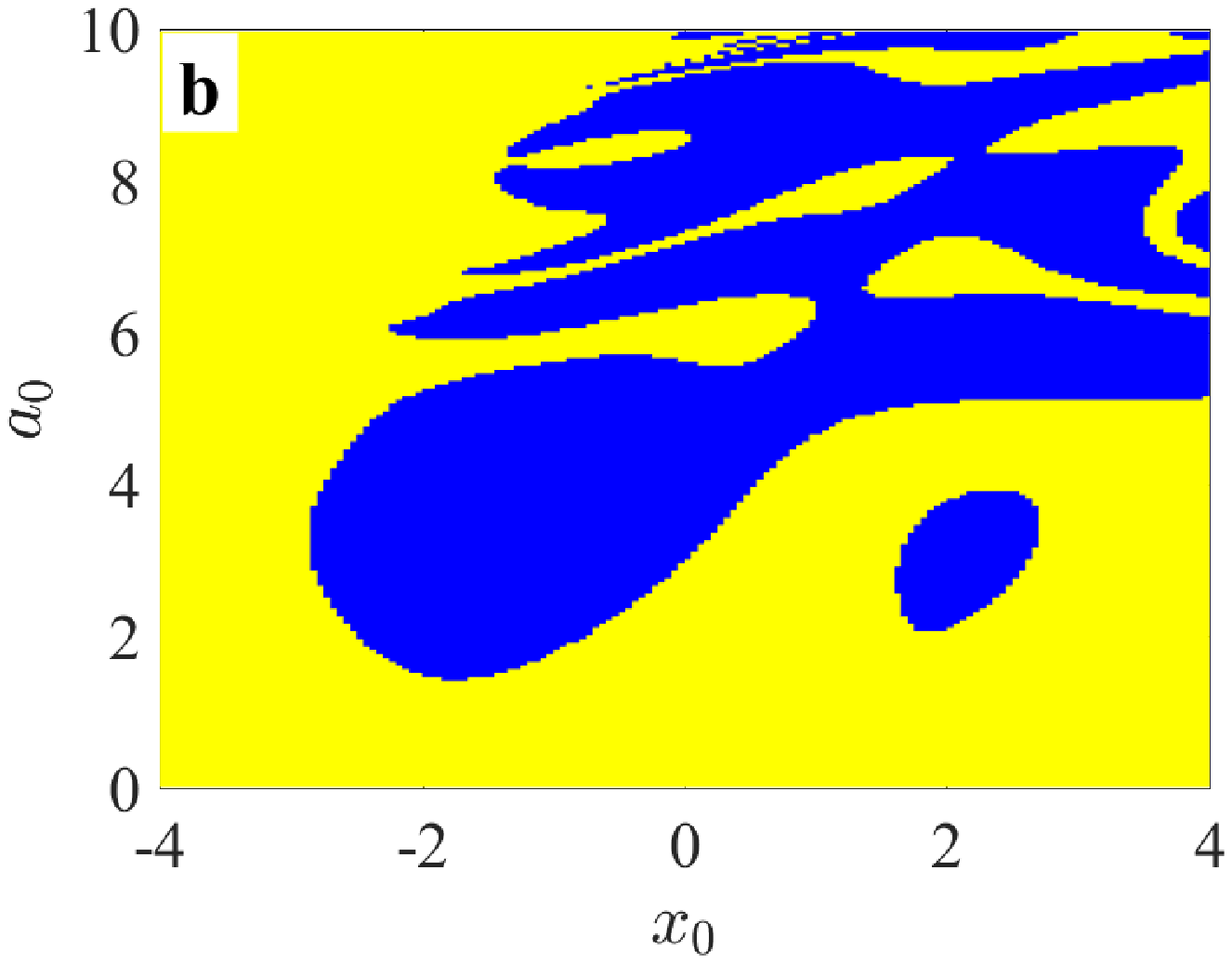}}     
     \subfigure
          {\includegraphics[width=0.40\columnwidth]{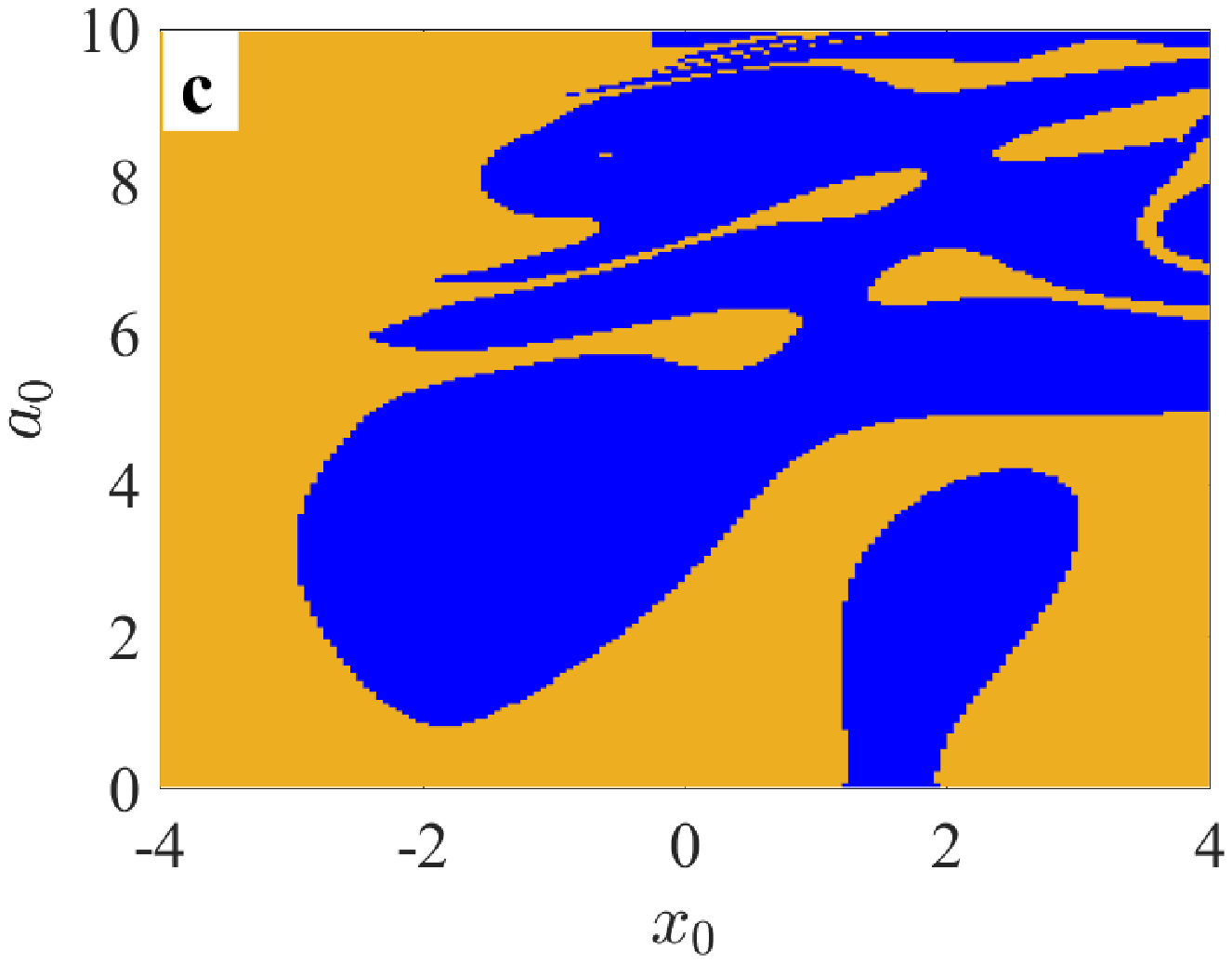}}
     \subfigure
          {\includegraphics[width=0.40\columnwidth]{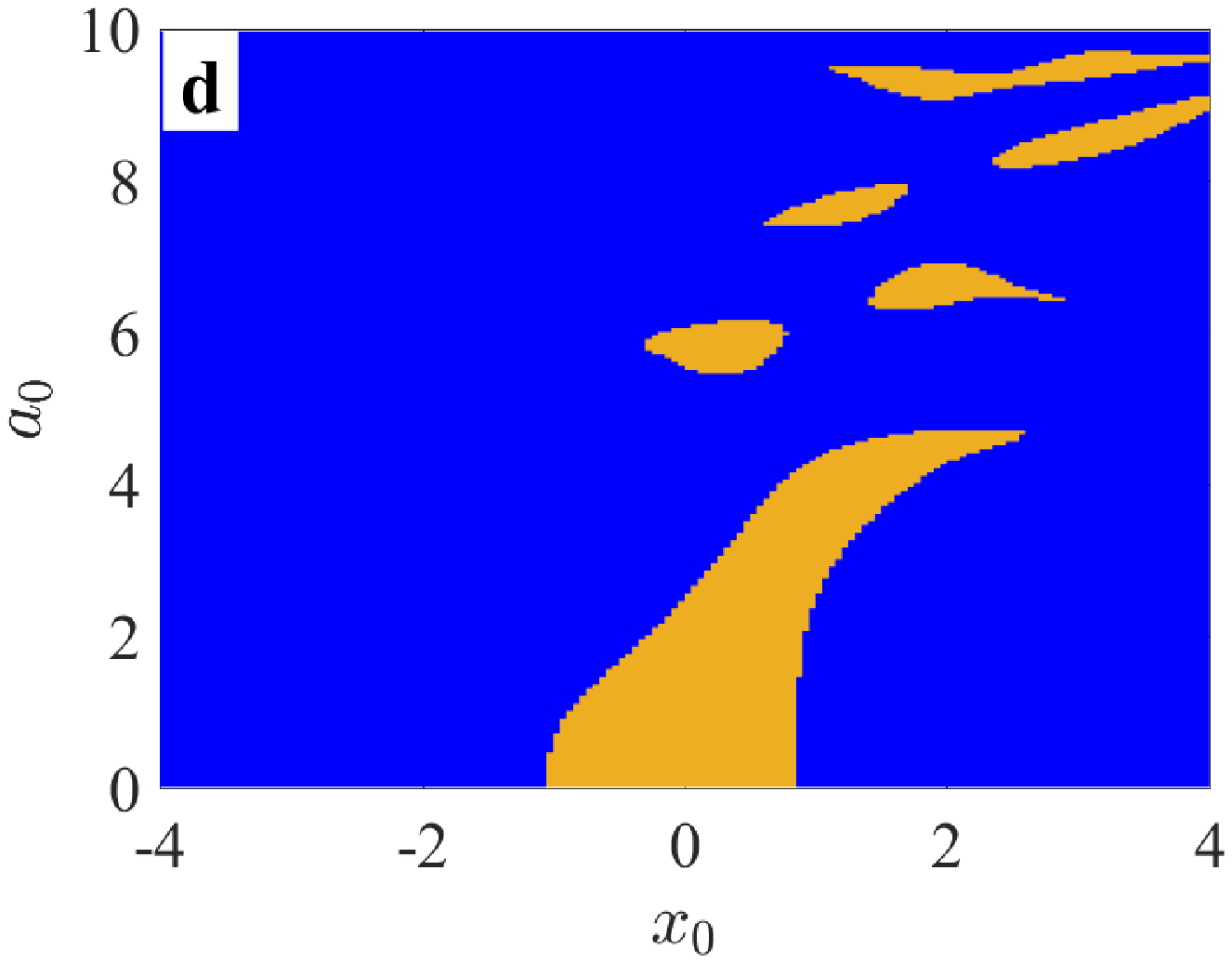}}
     \caption{(a)-(d) Basins of attraction in a the two-dimensional slice of the four-dimensional phase space $x_0=x(0), a_0=Re\{\tilde{a}(0)\}$, where the other initial conditions are $v(0)=0, Im\{\tilde{a}(0)\}=0$, for the values of detuning $\Delta$ depicted in the top subfigure: (a) $\Delta=0.56$, (b) $\Delta=0.68$, (c) $\Delta=0.71$, (d) $\Delta=0.74$. Other parameter values are as in Fig. 2(B). The basins of attraction of the fixed point (yellow), the large (blue) and the small (orange) limit cycle are shown. The small limit cycle emerges from the fixed point through a Hopf bifurcation and inherits its basin of attraction.}
\end{center}
\end{figure*}

\begin{figure*}\centering
\begin{center}
     \subfigure
          {\includegraphics[width=0.40\columnwidth]{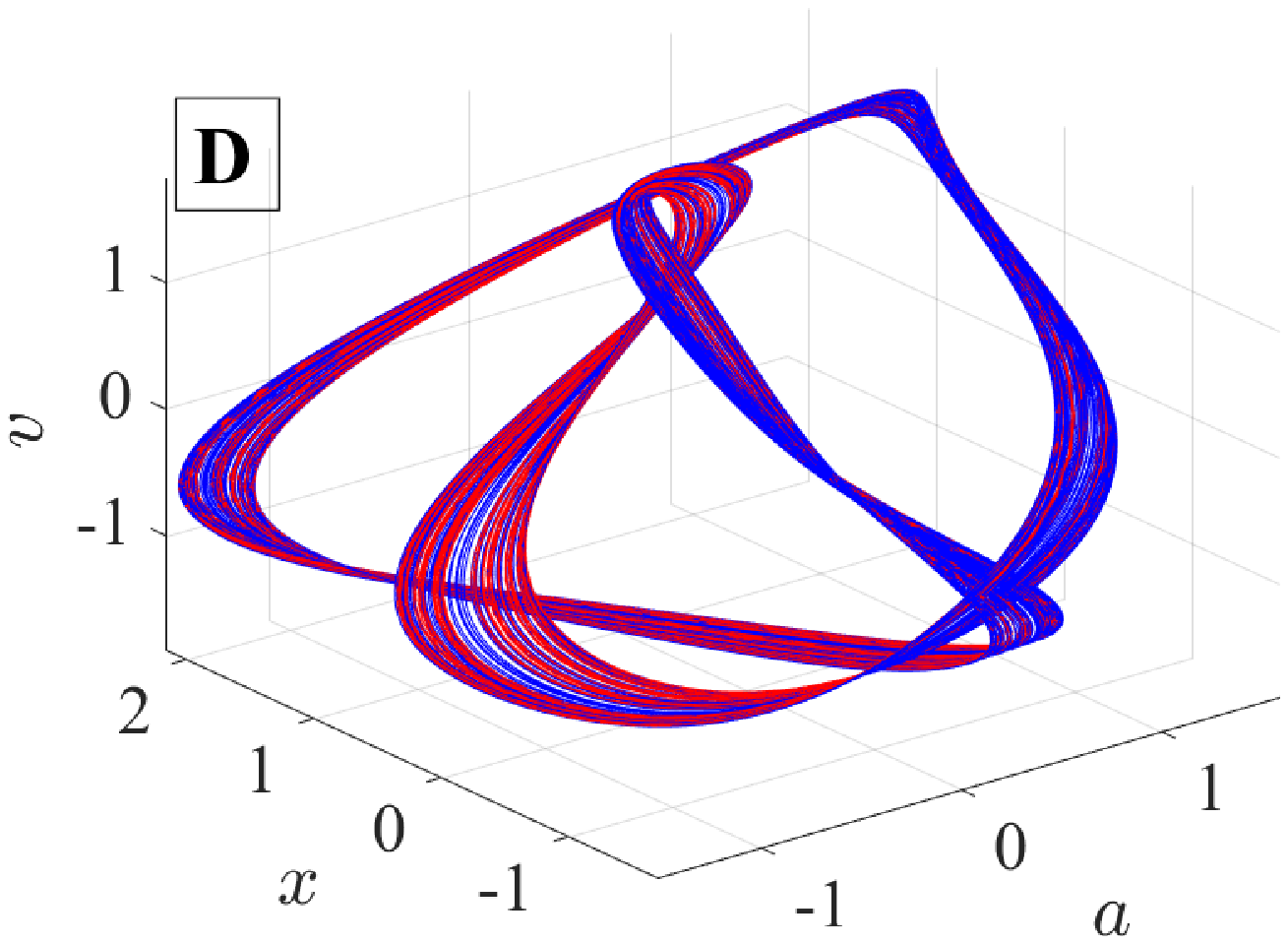}}
     \subfigure
          {\includegraphics[width=0.40\columnwidth]{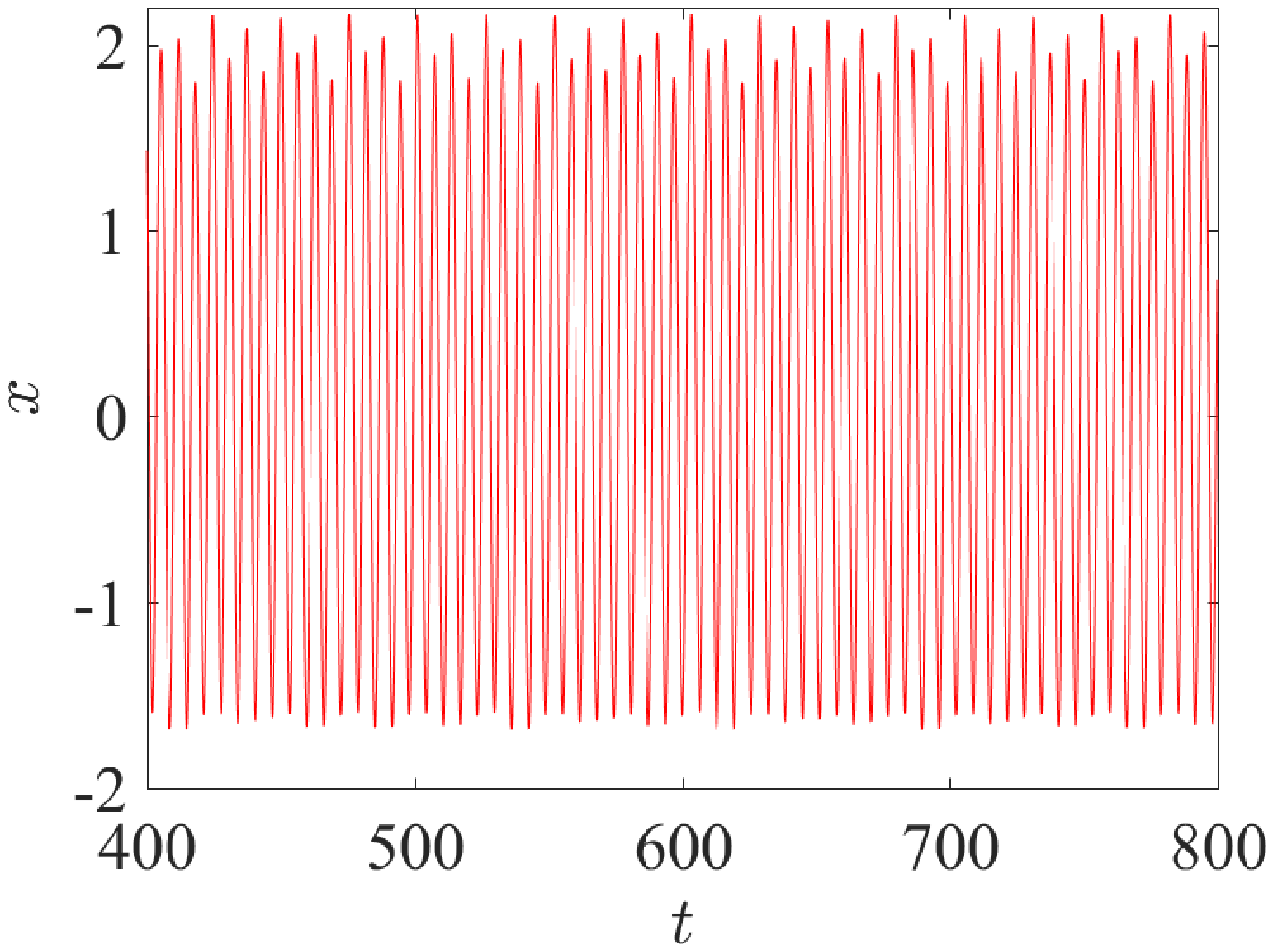}}  
     \subfigure
          {\includegraphics[width=0.40\columnwidth]{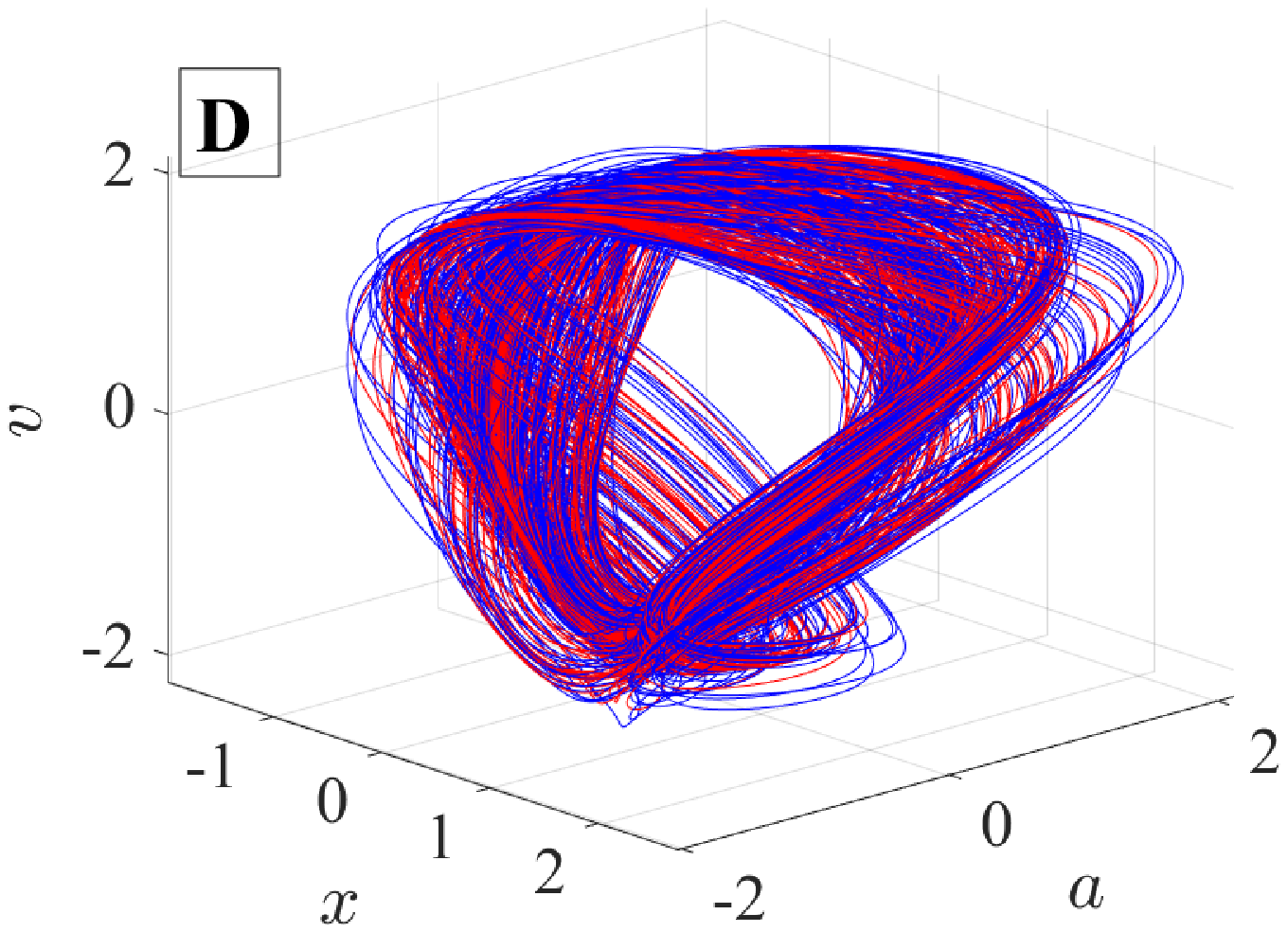}} 
     \subfigure
          {\includegraphics[width=0.40\columnwidth]{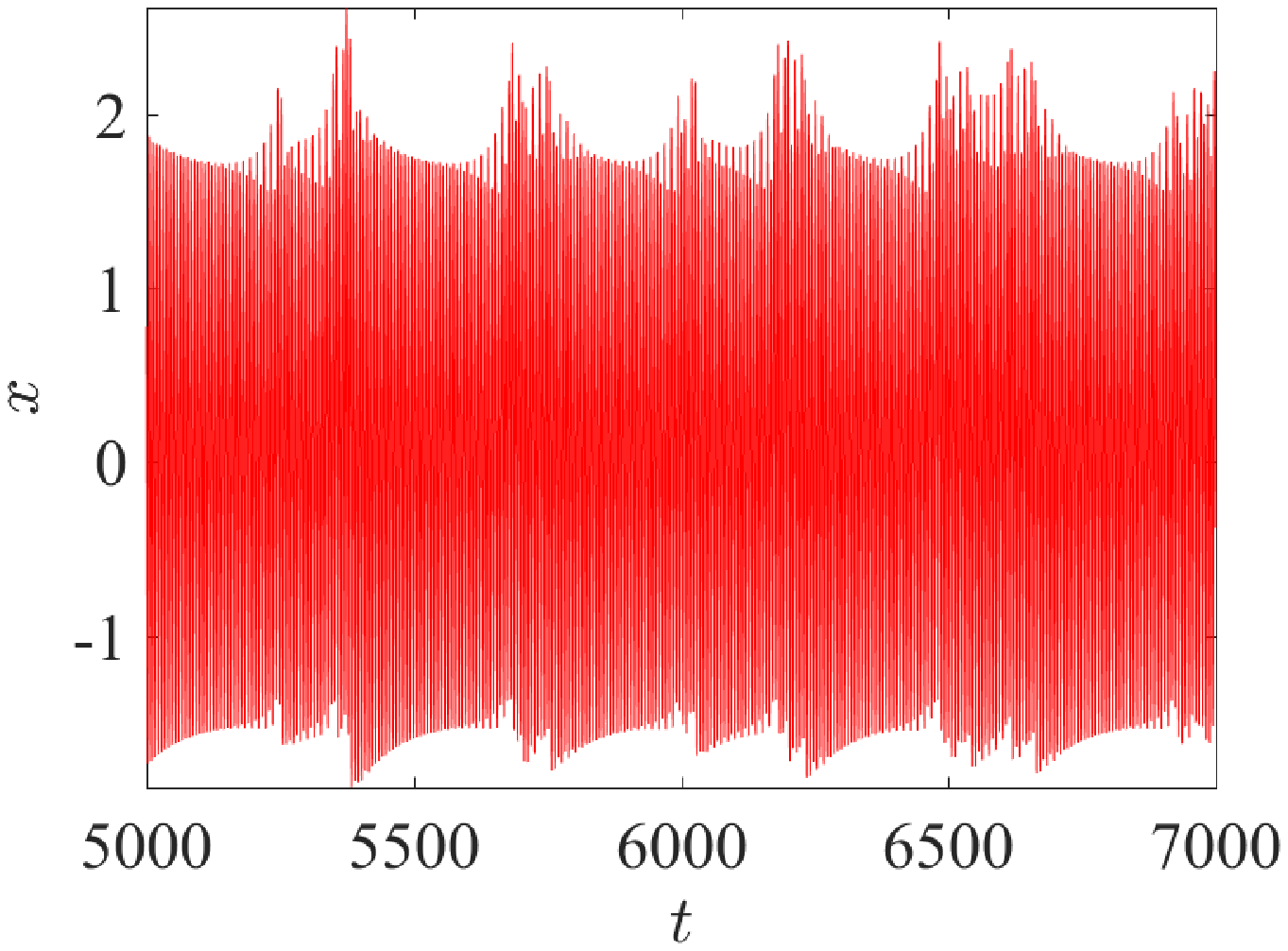}}
     \caption{(top) A quasi-periodically self-modulated oscillation forming a torus in the phase space, for $\Delta=0.2724$. (bottom) A chaotically self-modulated oscillation, for $\Delta=0.2300$. Other parameters: $\kappa=0.1, \gamma=0.01$ and $P=0.15$ (corresponding to the case (D) of Fig. 1 and Fig. 3). The absence of stable fixed points or limit cycles for these parameter values facilitates the accessibility of these oscillatory states from a large number of initial conditions. In the left panels, two different orbits are shown in red and blue.}
\end{center}
\end{figure*}


\clearpage 

\begin{thebibliography}{100}
\bibitem{RMP_14} M. Aspelmeyer, T.J. Kippenberg, and F. Marquardt, ``Cavity Optomechanics'', Rev. Mod. Phys. \textbf{86}, 1391-1452 (2014).

\bibitem{Metcalfe_14} M. Metcalfe, ``Applications of cavity optomechanics'', Appl. Phys. Rev. \textbf{1}, 031105 (2014).

\bibitem{Elste_09} F. Elste,S.M. Girvin, and A.A. Clerk, ``Quantum Noise Interference and Backaction Cooling in Cavity Nanomechanics'', Phys. Rev. Lett. \textbf{102}, 207209 (2009).

\bibitem{Painter_12} A.G. Krause, M. Winger, T.D. Blasius, Q. Lin, and O. Painter, ``A high-resolution microchip optomechanical accelerometer'', Nat. Photonics \textbf{6}, 768-772 (2012).

\bibitem{Yang_16a} Z.-P. Liu, J. Zhang, S.K. Ozdemir, B. Peng, H. Jing, X.-Y. Lu, C.-W. Li, L. Yang, F. Nori,and Y. Liu, ``Metrology with PT -Symmetric Cavities: Enhanced Sensitivity near the PT -Phase Transition'', Phys. Rev. Lett. \textbf{117}, 110802 (2016).

\bibitem{Harris_16} H. Xu, D. Mason, L. Jiang, and J.G.E. Harris, ``Topological energy transfer in an optomechanical system with exceptional points'', Nature \textbf{537}, 80-83 (2016).

\bibitem{Reed_17} A.P. Reed, K.H. Mayer, J.D. Teufel, L.D. Burkhart, W. Pfaff, M. Reagor, L. Sletten, X. Ma, R.J. Schoelkopf, E. Knill, and K.W. Lehnert, ``Faithful conversion of propagating quantum information to mechanical motion'', Nat. Physics \textbf{13}, 1163-1167 (2017).

\bibitem{Vahala_09} K. Vahala, M. Herrmann, S. Knunz, V. Batteiger, G. Saathoff, T.W. Hansch, and Th. Udem, ``A phonon laser'', Nat. Physics \textbf{5}, 682-686 (2009).

\bibitem{Yang_18} J. Zhang, B. Peng, S.K. Ozdemir, K. Pichler, D.O. Krimer, G. Zhao, F. Nori, Y. Liu, S. Rotter, and L. Yang, ``A phonon laser operating at an exceptional point'', Nat. Photonics\textbf{12}, 479-484 (2018).

\bibitem{Russell_14} A. Butsch, J.R. Koehler, R.E. Noskov, and P.St.J. Russell, ``CW-pumped single-pass frequency comb generation by resonant optomechanical nonlinearity in dual-nanoweb fiber'', Optica \textbf{1}, 158-164 (2014).

\bibitem{Russell_16} J.R. Koehler, R.E. Noskov, A.A. Sukhorukov, A. Butsch, D. Novoa, and P.St.J. Russell, ``Resolving the mystery of miliwatt-threshold opto-mechanical self-oscillation in dual-nanoweb fiber'', APL Photonics \textbf{1}, 056101 (2016).

\bibitem{Russell_17} J.R. Koehler, R.E. Noskov, A.A. Sukhorukov, D. Novoa, and P.St.J. Russell, ``Coherent control of flexural vibrations in dual-nanoweb fibers using phase-modulated two-frequency light'', Phys. Rev. A \textbf{96}, 063822 (2017).

\bibitem{Sukhorukov_18} R.E. Noskov, J.R. Koehler, and A.A. Sukhorukov, ``Interplay of Cascaded Raman- and Brilluin-like Scattering in Nanostructured Optical Waveguides'', ACS Photonics \textbf{5}, 1074-1083 (2018).


\bibitem{Dorsel_83} A. Dorsel, J.D. McCullen, P. Meystre, E. Vignes, and H. Walther, ``Optical Bistability and Mirror Confinement Induced by Radiation Pressure'', Phys. Rev. Lett. \textbf{51}, 1550-1553 (1983).

\bibitem{Mancini_94} S. Mancini and P. Tombesi, ``Quantum noise reduction by radiation pressure'', Phys. Rev. A \textbf{49}, 4055-4065 (1994).

\bibitem{Vahala_05a} T. Carmon, H. Rokhsari, L. Yang, T.J. Kippenberg, and K.J. Vahala, ``Temporal Behavior of Rdiation-Pressure-Induced Vibrations of an Optical Microcavity Phonon Mode'', Phys. Rev. Lett. \textbf{94}, 223902 (2005).

\bibitem{Vahala_05b} T.J. Kippenberg, H. Rokhsari, T. Carmon, A. Scherer, and K.J. Vahala, ``Analysis of Radiation-Pressure Indeuced Mechanical Oscillation of an Optical Microcavity'', Phys. Rev. Lett. \textbf{95}, 033901 (2005).

\bibitem{Marquardt_08a} C. Metzger, M. Ludwig, C. Neuenhahn, A. Ortlieb, I. Favero, K. Karrai, and F. Marquardt, ``Self-Induced Oscillations in an Optomechanical System Driven by Bolometric Backaction'', Phys. Rev. Lett. \textbf{101}, 133903 (2008).

\bibitem{Marquardt_08b} M. Ludwig, B. Kubala, and F. Marquardt, ``The optomechanical instability in the quantum regime'', New J. Phys. \textbf{10}, 095013 (2008).

\bibitem{Marquardt_14} N. Lorch, J. Qian, A. Clerk, F. Marquardt, and K. Hammerer, ``Laser Theory for Optomechanics: Limit Cycles in the Quantum Regime'', Phys. Rev. X \textbf{4}, 011015 (2014).

\bibitem{Painter_15} A.G. Krause, J.T. Hill, M. Ludwig, A.H. Safavi-Naeini, J. Chan, F. Marquardt, and O. Painter, ``Nonlinear Radiation Pressure Dynamics in an Optomechanical Crystal'', Phys. Rev. Lett. \textbf{15}, 233601 (2015).

\bibitem{Buters_15} F.M. Buters, H.J. Eerkens, K. Heeck, M.J. Weaver, B. Pepper, S. de Man, and D. Bouwmeester, ``Experimental exploration of the optomechanical attractor diagram and its dynamics'', Phys. Rev. A \textbf{92}, 013811 (2015).

\bibitem{Marquardt_06} F. Marquardt, J.G.E. Harris, and S.M. Girvin, ``Dynamical Multistability Induced by Radiation Pressure in High Finesse Micromechanical Optical Cavities'', Phys. Rev. Lett. \textbf{96}, 103901 (2006).

\bibitem{Fehske_16a} C. Schulz, A. Alvermann, L. Bakemeier, and H. Fehske, ``Optomechanical multistability in the quantum regime'', Europhys. Lett. \textbf{113}, 64002 (2016)

\bibitem{Vahala_07} T. Carmon, M.C. Cross, and K.J. Vahala, ``Chaotic Quivering of Micron-Scaled On-Chip Resonators Excited by Centrifugal Optical Pressure'', Phys. Rev. Lett. \textbf{98}, 167203 (2007).

\bibitem{Marino_11} F. Marino and F. Marin, ``Chaotically spiking attractors in suspended-mirror optical cavities'', Phys. Rev. E \textbf{83}, 015202(R) (2011).

\bibitem{Grebogi_14} G. Wang, L. Huang, Y.-C. Lai, and C. Grebogi, ``Nonlinear Dynamics and Quantum Entanglement in Optomechanical Systems'', Phys. Rev. Lett. \textbf{112}, 110406 (2014).

\bibitem{Yang_16b} F. Monifi, J. Zhang, S.K. Ozdemir, B. Peng, Y. Liu, F. Bo, F. Nori, and L. Yang, ``Optomechanically induced stochastic resonance and chaor transfer between optical fields'', Nat. Photonics \textbf{10}, 399-405 (2016).

\bibitem{Liu_17} J. Wu, S.-W. Huang, Y. Huang, H. Zhou, J. Yang, J.-M. Liu, M.Yu, G. Lo, D.-L. Kwong, S. Duan, and C.W. Wong, ``Mesoscopic chaos mediated by Drude electron-hole plasma in silicon optomechanical oscillators'', Nat. Commun. \textbf{8}, 15570 (2017).

\bibitem{Marquardt_20} T. Figueiredo Roque, F. Marquardt, and O.M. Yevtushenko, ``Nonlinear dynamics of weakly dissipative optomechanical systems'', New J. Phys. \textbf{22}, 013049 (2020).

\bibitem{Feshke_15} L. Bakemeier, A. Alvermann, and H. Fehske, ``Route to Chaos in Optomechanics'', Phys. Rev. Lett. \textbf{114}, 013601 (2015).

\bibitem{Wang_16} M. Wang, X.-Y. Lu, J.-Y. Ma, H. Xiong, L.-G. Si, and Y. Wu, ''Controllable chaos in hybrid electro-optomechanical systems``, Sci. Rep. \textbf{6}, 22705 (2016).

\bibitem{Urrios_17} D. Navarro-Urrios, N.E. Capuj, M.F. Colombano, P.D. Garcia, M. Sledzinska, F. Alzina, A. Griol, A. Martinez, and C.M. Sotomayor-Torres, ''Nonlinear dynamics and chaos in an optomechanical beam``, Nat. Commun. \textbf{8}, 14965 (2017).

\bibitem{Lu_20} D.-W. Zhang, C. You, and X.-Y. Lu, ''Intemittent chaos in cavity optomechanics``, Phys. Rev. A \textbf{101}, 053851 (2020).
 
\bibitem{Kingni_20} S.T. Kingni, C. Tchodimou, D.P. Foulla, P. Djorwe, and S.G. Nana Engo, ''Antimonotonicity, coexisting attractors and bursting oscillations in optomechanical system: Analysis and electronic implementation``, Eur. Phys. J. Special Topics \textbf{229}, 1117-1132 (2020).

\bibitem{Marquardt_11} G. Heinrich, M. Ludwig, J. Qian, B. Kubala, and F. Marquardt, ''Collective Dynamics in Optomechanical Arrays``, Phys. Rev. Lett. \textbf{107}, 043603 (2011).

\bibitem{Zhang_12} M. Xhang, G.S. Wiederhecker, S. Manipatruni, A. Barnard, P. McEuen, and M. Lipson, ''Synchronization of Micromechanical Oscillators Using Light``, Phys. Rev. Lett. \textbf{109}, 233906 (2012).

\bibitem{Milburn_12} C.A. Holmes, C.P Meaney, and M.J. Milburn, ''Synchronization of many nanomechanical resonators coupled via a common cavity field``, Phys. Rev. E \textbf{85}, 066203 (2012).

\bibitem{Marquardt_13} M. Bagheri, M. Poot, L. Fan, F. Marquardt, and H.X. Tang, ''Photonic Cavity Synchronization of Nanomechanical Oscillators``, Phys. Rev. Lett. \textbf{111}, 213902 (2013).

\bibitem{Marquardt_16} T. Weiss, A. Kronwald, and F. Marquardt, ''Noise-induced transitions in optomechanical synchronization``, New J. Phys. \textbf{18}, 013043 (2016).

\bibitem{ElGanainy_16} D.W. Schonleber, A. Eisfeld, and R. El-Ganainy, ''Optomechanical interactions in non-Hermitian photonic molecules``, New J. Phys. \textbf{18}, 045014 (2016).

\bibitem{Fehske_16b} C. Wurl, A. Alvermann, and H. Fehske, ''Symmetry-breaking oscillations in membrane optomechanics``, Phys. Rev. A \textbf{94}, 063860 (2016). 

\bibitem{Alu_17} M.-A. Miri, E. Verhagen, and A. Alu, ''Optimechanically induced spontaneous symmetry breaking``, Phys. Rev. A \textbf{95}, 053822 (2017). 

\bibitem{Nori_19} N. Yang, A. Miranowicz, Y.-C. Liu, K. Xia, and F. Nori, ''Chaotic synchronization of two optical cavity modes in optomechanichal systems``, Sci. Rep. \textbf{9}, 15874 (2019).

\bibitem{Vitali_20} W. Li, P. Piergentili, J. Li, S. Zippilli, R. Natali, N. Malossi, G. Di Giuseppe, and D. Vitali, ''Noise robustness of synchronization of two nanomechanical resonators coupled to the same cavity field``, Phys. Rev. A \textbf{101}, 013802 (2020).

\bibitem{Kovanis_95a} V. Kovanis, A. Gavrielides, T.B. Simpson, and J.M. Liu, ''Instabilities and chaos in optically injected semiconductor lasers``, Appl. Phys. Lett. \textbf{67} 2780-2782 (1995). 

\bibitem{Kovanis_95b}  T.B. Simpson, J.M. Liu, A. Gavrielides, V. Kovanis, and P.M. Alsing, ''Period-doubling cascades and chaos in a semiconductor laser with optical injection``, Phys. Rev. A \textbf{51}, 4181-4185 (1995).

\bibitem{Kovanis_97} A. Gavrielides, V. Kovanis, P.M. Varangis, T. Erneux, and G. Lythe, ''Coexisting periodic attractors in injection-locked diode lasers``, Quantum Semiclass. Opt. \textbf{9}, 785-796 (1997).

\bibitem{Kominis_17} Y. Kominis, V. Kovanis, and T. Bountis, ''Controllable asymmetric phase-locked states of the fundamental active photonic dimer``, Phys. Rev. A \textbf{96}, 043836 (2017).

\bibitem{Kominis_20} Y. Kominis, A. Bountis, and V. Kovanis, ''Radically tunable ultrafast photonic oscillators via differential pumping``, J. Appl. Phys. \textbf{127}, 083103 (2020).

\bibitem{Kuznetsov} Y. Kuznetsov, Elements of Applied Bifurcation Theory, Applied Mathematical Sciences (Springer, Berlin, 2004).

\bibitem{MATCONT} A. Dhooge, and W. Govaerts, Yu.A. Kuznetsov, H.G.E. Meijer, and B. Sautois, ''New features of the software MatCont for bifurcation analysis of dynamical systems``, Math. Comput. Model Dyn. Syst. \textbf{14}, 147-175 (2008).



\end{thebibliography}
\end{document}